\documentclass[twocolumn]{aastex631}

\usepackage{amsmath}
\usepackage{amsfonts}
\usepackage{amssymb}
\usepackage{graphicx}
\usepackage{hyperref}
\usepackage{xcolor}
\usepackage{float}
\usepackage{soul}
\usepackage[normalem]{ulem}
\usepackage{subfigure}

\def\kms{$\mathrm {km s}^{-1}$}
\def\WmT{$\mathrm{W m}^{-2}\mathrm{T}^{-4}$}
\def\apjl{ApJ}                
\def\apjs{ApJS}               
\def\aap{A\&A}                

\newcommand{\A}{\textsc{a}}
\newcommand{\B}{\textsc{b}}
\newcommand{\C}{\textsc{c}}
\newcommand{\D}{\textsc{d}}

\newcommand{\T}{\text}

\graphicspath{{./}{figures/}}

\begin{document}

\title{The impact of binaries on the dynamical mass estimate of dwarf galaxies}

\author{Camilla Pianta}
\affiliation{Dipartimento di Fisica e Astronomia Galileo Galilei, Università degli Studi di Padova, Vicolo Osservatorio 3, I-35122, Padova, Italy}

\author{Roberto Capuzzo-Dolcetta}
\affiliation{Dipartimento di Fisica, Università di Roma La Sapienza, Piazzale Aldo Moro 5, I-00185, Roma, Italy}

\author{Giovanni Carraro}
\affiliation{Dipartimento di Fisica e Astronomia Galileo Galilei, Università degli Studi di Padova, Vicolo Osservatorio 3, I-35122, Padova, Italy}

\begin{abstract}
Binary stars are recognized to be important in driving the dynamical evolution of stellar systems and also in determining some of their observational features. In this study, we explore the role that binary stars have in modulating the estimates of the velocity dispersion of stellar systems. To this aim, we developed a tool which allows to investigate the dependence of synthetic velocity dispersion on a number of crucial quantities characterizing the binary content: binary fraction and the distributions of their mass ratio, eccentricity and semi-major axis. As an application, we evaluate the impact that binary stars have on the estimation of the dynamical mass of dwarf spheroidal and ultra-faint dwarf galaxies, finding that it can be particularly relevant, especially for low mass and low density systems. These results bear profound implications for the interpretation of the measured velocity dispersion in such systems, since it weakens or relieves the claim for the need of large amounts of dark matter.

\end{abstract}


\section{Introduction}

The interest in Local Group dwarf spheroidal galaxies (dSph) has been growing over the last decades due to their large mass-to-light ratios, as obtained by analyzing their stellar kinematics (\citeauthor{2011MNRAS.411.2118A}, \citeyear{2011MNRAS.411.2118A}). These objects, located at least 70 kpc away and composed by old and metal-poor stars, do not show any clear rotation, so they are not centrifugally supported: this implies that any attempt of estimating their masses must involve the spectroscopic measurement of their velocity dispersion (\citeauthor{1993AJ....105..510M}, \citeyear{1993AJ....105..510M}). In particular, several studies conducted on classical dSphs in the Milky Way (MW) halo (i.e., Fornax, Sculptor, Ursa Minor I, Draco, Leo I, Leo II, Sagittarius, Sextans and LGS 3) pointed out that the observed velocity dispersion, $\sigma_{obs}$, is significantly inflated with respect to the expected value, which would be of the order of $1-3$ \kms~, if globular clusters' (GC) kinematic properties were scaled by the structural parameters of dSphs (\citeauthor{1997ASPC..116..259M}, \citeyear{1997ASPC..116..259M}). Additional research (\citeauthor{2007ApJ...670..313S}, \citeyear{2007ApJ...670..313S}) corroborates this result through the detection of values between $3.3-7.6$ \kms~ for the observed velocity dispersion, thus challenging the claim about the existence of dSphs having $\sigma_{obs} \le$ 7 \kms (\citeauthor{2008IAUS..244...44W}, \citeyear{2008IAUS..244...44W}), and consequently revising the mass limit for such systems.\\
Different scenarios to account for the notably large velocity dispersion of dSphs have been proposed: one (see for instance \citeauthor{1988IAUS..130..409A}, \citeyear{1988IAUS..130..409A}) asserts the presence of a considerable amount of dark matter (DM), while another suggests that dSphs are not in virial equilibrium, but actually ongoing tidal disruption.
However, the role of galactic tides has been weakened by considerations based on the luminosity-metallicity relationship (\citeauthor{2008ApJ...685L..43K}, \citeyear{2008ApJ...685L..43K}), and on a missing unambiguous identification of both kinematic outliers in the observed stellar samples and stream motions for dSphs in the proximity of the MW. By way of example, no evidence of either tidal tails or induced rotation was found in Segue I by \citeauthor{2009ApJ...692.1464G} (\citeyear{2009ApJ...692.1464G}), who rejected the hypothesis of such a system being a GC once associated with the Sagittarius (Sgr) stream (\citeauthor{2007ApJ...654..897B}, \citeyear{2007ApJ...654..897B}).\\
A further possibility is that the high values of the observed velocity dispersion are due to binary orbital motion. \textit{De facto}, while the tidal stripping scenario has been refuted, the role of binaries has been, and still is, object of investigation. According to \citeauthor{1997ASPC..116..259M} (\citeyear{1997ASPC..116..259M}), the presence of unresolved binary stars, independently of their fraction, is unlikely to be fully responsible for the inflation of $\sigma_{obs}$ in classical dSphs, which are hereby regarded as DM-dominated systems. Instead, their impact may be non-negligible in the case of ultra-faint dwarf (UFD) galaxies, i.e., the low-luminosity counterparts of classical dSphs (\citeauthor{2010ApJ...722L.209M}, \citeyear{2010ApJ...722L.209M}; \citeauthor{2018AJ....156..257S}, \citeyear{2018AJ....156..257S}). In spite of the fact that the sample of the examined UFDs has been moderately enlarged lately (\citeauthor{2018A&A...620A.155M}, \citeyear{2018A&A...620A.155M}), the small number statistics and the lack of appropriate multi-epoch observations remain a major problem in giving a safe estimate of their binary fraction and period distribution (\citeauthor{2010ApJ...722L.209M}, \citeyear{2010ApJ...722L.209M}).  To this end, it is worth mentioning the case of Segue II, whose velocity dispersion inflation has been extensively debated (\citeauthor{2009MNRAS.397.1748B}, \citeyear{2009MNRAS.397.1748B}; \citeauthor{2013ApJ...770...16K}, \citeyear{2013ApJ...770...16K}). So, unfortunately, only in quite a few instances the available spectroscopic data allow to constrain the binary fraction (e.g., for the UFD galaxy Reticulum II  (\citeauthor{2019MNRAS.487.2961M}, \citeyear{2019MNRAS.487.2961M})). For this reason, a modeling approach consisting in Monte Carlo simulations and Bayesian analysis has been frequently adopted. Up to now, most models have been trying to reproduce the observed velocity dispersion of classical dSphs by varying both the binary fraction and the binary orbital parameters, and have then compared the results to spectroscopic data in order to make estimates about the extent of the binary contribution to $\sigma_{obs}$ in UFDs  (\citeauthor{2017AJ....153..254S}, \citeyear{2017AJ....153..254S}; \citeauthor{2018A&A...620A.155M}, \citeyear{2018A&A...620A.155M}). Still, the assumptions on the orbital parameter distributions, especially related to periods and semi-major axes, are an actual limitation in this context: hence the desire of a theoretical model to make as reliable and general as possible inferences about the binary population of such systems.\\
Bearing in mind that the hypothesis of both dSphs and UFDs to be DM-dominated is currently the most supported one, we present in this paper a parametric study to explore the effects of the orbital parameters choice at varying binary fraction on the observed velocity dispersion of such galaxies. The ultimate purpose of this work is, therefore, investigating the impact of binary stars on the determination of the dynamical mass in the faintest MW satellites, with particular reference to that of \citeauthor{2020ApJ...896..152R} (\citeyear{2020ApJ...896..152R}) on OCs as far as the methodology to calculate the velocity dispersion is concerned.\\

The paper is organized as follows: in Sect. \ref{meth} we introduce and explain the methodology adopted, and describe the characteristics of our set of simulations in accordance with various choices for the binary population; in Sect. \ref{res} we critically expose our results; finally, in Sect. \ref{conc} we extract the main conclusions of our work.
\vfill\eject

\section{Methodology} \label{meth}

We built up a parametric model assuming as star density distribution that of a Plummer sphere of scale radius $R$ and total mass $M$, according to the law

\begin{equation}
    \rho(r) = \frac{3M}{4\pi R^3} \Bigg[1+\bigg(\frac{r}{R}\bigg)^2\Bigg]^{-\frac{5}{2}}.
\end{equation}

We reproduced both a standard dSph galaxy with a scale radius $R=3$ kpc, a total stellar mass $M=10^{7}$ M$_\odot$ (\citeauthor{2008Natur.454.1096S}, \citeyear{2008Natur.454.1096S}) and an age of 13 Gyrs, and a UFD of the same age, with a scale radius $R=50$ pc and a total stellar mass $M=5 \times 10^{4}$ M$_\odot$.\\

By means of these structural parameters, we computed the half-mass relaxation time (\citeauthor{2018ApJ...855...87M}, \citeyear{2018ApJ...855...87M})

\begin{equation} \label{t_rh}
    t_{rh} = \frac{\gamma N}{\ln\Lambda} \sqrt{\frac{r_h^3}{GM}} \simeq \frac{\gamma N}{\ln(\lambda N)} \sqrt{\frac{(1.3R)^3}{GM}},
\end{equation}
\\
with $\gamma =$ 0.138, $\lambda \approx$ 0.11, $G$ gravitational constant, $N$ total number of stars, $M$ total mass and $r_h=1.3R$ half-mass radius of the system. See Tab. \ref{structural_parameters} for a summary of the main features of the simulated galaxies.

\begin{deluxetable*}{c c c c c c c c c c c}[t]
\tablecaption{Structural parameters of the simulated galaxies.} 
\label{structural_parameters}
\tablehead{
\colhead{Object} & \colhead{$R$} & \colhead{$M$} & \colhead{$t_{rh}$} & \colhead{$X$} & \colhead{$Y$} & \colhead{$Z$} & \colhead{Age} & \colhead{$L_{bol}$} & \colhead{$L_V$} & \colhead{$L_B$} \\
\colhead{} & \colhead{(pc)} & \colhead{(M$_\odot$)} & \colhead{(Gyr)} & \colhead{} & \colhead{} & \colhead{} & \colhead{(Gyr)} & \colhead{(L$_\odot$)} & \colhead{(L$_{V,\odot}$)} & \colhead{(L$_{B,\odot}$)} \\
}
\startdata
dSph & $3 \times 10^3$ & $10^7$ & $1.79 \times 10^5$ & 0.747 & 0.252 & 0.001 & 13 & $1.35 \times 10^8$ & $1.38 \times 10^7$ & $1.67 \times 10^7$\\
UFD & 50 & $5 \times 10^4$ & 43.07 & 0.747 & 0.252 & 0.001 & 13 & $6.72 \times 10^5$ & $6.88 \times 10^4$ & $8.37 \times 10^4$ \\
\enddata
\end{deluxetable*}

The discrete stellar mass population is generated by sampling the Kroupa IMF (\citeauthor{2001MNRAS.322..231K}, \citeyear{2001MNRAS.322..231K}) in the interval [0.1, 50] M$_\odot$, i.e.,

\begin{equation}
    f(m)\propto m^{-\alpha}, \, \textrm{with} \,  \begin{cases}
    \alpha=1.3, ~\textrm{for} ~0.1\leq m/M_\odot<0.5,\\
    \alpha=2.3,~\textrm{for} ~0.5\leq m/M_\odot\leq 50,\end{cases}
\end{equation}
\\
\noindent
where the normalization constants are such to give a matching of the two power laws passing from a mass interval to the adjacent. 
The average star mass results $\langle m\rangle=0.61$ M$_\odot$. 

In the total number of stars in the system, $N$, we included also binaries: the binary fraction is defined as $f_b=N_b/N$, where $N_b$ is the number of stellar pairs (i.e., binaries).
Consequently, $N=N_s+2N_b$, where $N_s$ is the number of single stars.\\
Our standard modeling of the binary star population consists in a selection of $N_s$ values from a given sample, and in a random pairing of the other $2N_b$ ones, with the most massive member designated as the primary star ($m_1$) and the lightest as the secondary ($m_2$). Of course, $m_b=m_1+m_2$ yields the mass of the binary. As an alternative to this method, we adopted a power-law mass-ratio distribution $p(q) \propto q^{-0.4}$ (\citeauthor{2008A&A...480..103K}, \citeyear{2008A&A...480..103K}), where $q=m_{2}/m_{1}$, with extremes $q_{min} =$ 0.1 and $q_{max} =$ 1 (\citeauthor{2020ApJ...896..152R}, \citeyear{2020ApJ...896..152R}), to couple binary components in the case of the UFD model.\\
Upon the assumptions made for the age of the system and its chemical composition, we assigned to every star an evolutionary stage which characterizes it as Main-Sequence, Sub Giant, Red Giant, Asymptotic Giant, Horizontal Branch (all luminous objects), or as White Dwarf (WD), Neutron Star (NS) or Black Hole (BH) dark remnant. 
Note that the baryonic stellar `dark' component (WDs+NSs+BHs) in our model comprises a fraction of about $54\%$ of the total stellar mass.

The binary orbital parameters (Tab. \ref{simulations_summary}) are defined by the choice of a thermal eccentricity distribution $k(e)=2e$ (\citeauthor{1919MNRAS..79..408J} \citeyear{1919MNRAS..79..408J}), in the range $0\leq e \leq 1$, and a logarithmically flat semi-major axis distribution $g(a)\propto 1/a$ in the interval $a_{min}\leq a\leq a_{max}$, corresponding (at fixed $m_1+m_2$) to the period distribution

\begin{equation}
    h(P) = \frac{g(a)}{\frac{dP}{da}},
\end{equation}

which, once $a$ is expressed in terms of $P$ by the Kepler's third law, gives

\begin{equation}
    h(P) \propto\frac{1}{P},
\end{equation}

with $7 \times 10^{-2} \leq P~(\textrm{days}) \leq 6 \times 10^6$, values in good agreement with \citeauthor{1991A&A...248..485D} (\citeyear{1991A&A...248..485D}) and \citeauthor{2001ApJ...555..945K} (\citeyear{2001ApJ...555..945K}).

\begin{deluxetable}{c c c c}[h]
\tablecaption{Ranges of variation of parameters characterizing the binary populations.}
\label{simulations_summary}
\tablehead{
\colhead{$f_b$} & \colhead{$a_{min}$} & \colhead{$a_{max}$} &  \colhead{$e$}\\
\colhead{} & \colhead{(AU)} & \colhead{(AU)} &  \colhead{}
}
\startdata
0.05--0.4 & 0.01--1 & 50--400 & 0--1 \\
\enddata
\end{deluxetable}

In order to assess the impact of binary orbital motion on the observed velocity dispersion, we investigated how the variation of binary orbital parameters affects such a quantity. Unsurprisingly, the semi-major axis distribution turns out to be the most relevant within this framework, since the shrinking of the distance between binary components has a major effect on the estimate of the velocity dispersion. Hence, we first varied the upper boundary $a_{max}$ in the set of values [50, 100, 200, 300, 400] AU by keeping fixed the lower one, $a_{min}$, at 0.2 AU, and then we did the opposite, i.e., we selected the lower boundary in the range of values [0.01, 0.02, 0.03, 0.05, 0.08, 0.1, 0.2, 0.4, 0.6, 1] AU and settled the upper one to 100 AU. We repeated such a procedure for different binary fractions, going from 0.05 to 0.4 in steps of 0.05, and ran a hundred simulations for each one after having selected the semi-major axis distribution's extremes; in the end, we averaged data from each set to have a more robust statistical significance of the output.\\
In addition to this, we ran another set of simulations for both our model galaxies by accounting for the possible occurrence of the Roche Lobe Overflow (RLOF) phenomenon between close binary components. We deemed as undergoing RLOF merger all pairs whose components' stellar radii exceed the respective Roche-Lobe radii. Into specifics, we calculated the former as photospheric radii

\begin{equation}
    R_{phot} = \sqrt{\frac{L}{4\pi k_B T_{eff}^4}}
\end{equation}

where $L$ represents the stellar luminosity, $T_{eff}$ the effective temperature and $k_B=5.670 \times 10^{(-8)}$ \WmT ~the Boltzmann constant, and the latter through the Eggleton's formula (\citeauthor{1983ApJ...268..368E}, \citeyear{1983ApJ...268..368E})

\begin{equation}
    R_L = \frac{0.49q^{\frac{2}{3}}}{0.6q^{\frac{2}{3}} + \ln{\bigl(1+q^{\frac{1}{3}}}\bigr)}r_p.
\end{equation}

scaled by the pericenter distance $r_p = a(1-e)$ according to the prescription of \citeauthor{2007ApJ...660.1624S} (\citeyear{2007ApJ...660.1624S}).Hereby, binaries experiencing RLOF in both their components are considered as single (merged) objects and contribute to the observed velocity dispersion with their center of mass velocity (Eq. \ref{sigma_tot_RLOF}, Eq. \ref{sigma_tot_lum_RLOF}); on the other hand, binaries characterized by only one component overfilling its Roche Lobe cannot be regarded as such because the outcome of the mass transfer is actually uncertain.\\
Yet, a primary overflowing its Roche lobe triggers a sudden mass loss, which would cause a rapid modification of the host binary structure, concerning mainly its luminosity and effective temperature: in particular, the luminosity decline may be such prominent to make the binary slip out of a magnitude-limited stellar sample. \textit{Ergo}, the assumption that a quick merger between binary components happens when their Roche lobes touch is not fully correct. Mindful of this, we took a conservative approach by assigning to each merged binary a velocity equal to its previous center of mass one, given that following the time evolution of the simulated binary population, rather than examining its present configuration, would have not only introduced further complications and approximations in our analysis, but also rendered our results less accurate.\\
On top of that, we performed a luminosity cut-off consisting in the removal of all stars with luminosity below the turn-off (TO) level, condition given by $L_1 + L_2 < L_{TO}$ as for binaries, with the aim of mimicking a realistic observational situation. We stress that this operation is actually meaningful only for dSphs, whose velocity dispersion is typically derived from the fiber-fed multi-object spectroscopy of individual sources: therefore, only stars brighter than a certain threshold, to second of the instrumental set-up, can be fruitfully used. In the case of an UFD, instead, the velocity dispersion is routinely obtained from integrated single slit spectroscopy, which collects all the underlying light.
\\
We computed the observed velocity dispersion by considering binaries as unresolved (Eq. \ref{sigma_tot}, Eq. \ref{sigma_tot_lum}).
In contrast, $\sigma_{s,b}$ (Eq. \ref{sigma_sb}), $\sigma_s$ (Eq. \ref{sigma_s}), and $\sigma_{s,lum}$ (Eq. \ref{sigma_s_lum}) are not affected by the binary orbital motion for they represent, respectively, the velocity dispersion of single stars and binary centers of mass ($\sigma_{s,b}$), and the velocity dispersion of single stars only, where $\sigma_{s,lum}$ is a luminosity averaged value. As such, they do not depend on the variation of the binary semi-major axes and eccentricity, nor on the binary fraction. For this reason, we let $\sigma_{s} \equiv \sigma_0$ as identification of the intrinsic velocity dispersion, i.e., the velocity dispersion deriving from the structural parameters of the galaxy, defined in the assumption of global virial equilibrium by the equation

\begin{equation} \label{sigma_int}
    \sigma_{int} = \sqrt{\frac{|\Omega|}{M}},
\end{equation}

where $\Omega$ is the gravitational potential energy. In the following, we will refer to $\sigma_{int}$ as $\sigma_0$.\\

Moreover, we adopted as {\it reference} model for the binary population in our study that corresponding to (i) a random pairing of their masses with (ii) a logarithmically flat semi-major axis distribution in the interval [0.2, 100] AU and (iii) a thermal eccentricity distribution. Finally, we chose $\sigma_{tot}$ (Eq. \ref{sigma_tot}) as the observed velocity dispersion to determine the virial mass of our mocked galaxies.

\section{Results and discussion} \label{res}

In this section we provide a detailed look at the results of our simulations, highlighting how the assumptions made on the binary population reflect upon our model galaxies' dynamical mass estimate.

\subsection{Variation of binary orbital parameters}

As a general, preliminary, consideration, we point out that the observed velocity dispersion in a star system hosting a given set of binaries in a fraction $f_b$ can be represented as a linear combination of the two (single star and unresolved binary) contributions:

\begin{equation}
\sigma^2_{obs} = (1-f_b)\sigma_s^2 +f_b\sigma_{b}^2.
\end{equation}

Being $\sigma_s^2 \propto |\Omega|/M \propto M$ (Eq. \ref{sigma_int}), and $\sigma_b^2$ independent of $M$, it is clear that, once a specific binary population is generated, the action of binaries is as more relevant as lighter the system is, even in the case of small binary fractions. So it is natural to expect a major enhancement of the output velocity dispersion in UFDs than in dSphs: this is indeed confirmed by our thorough modelization.\\
In the hypothesis of virialized (i.e., stationary) systems, we can infer the relative variation of the predicted virial mass with respect to the real one via the expression

\begin{equation}
    \frac{\Delta M}{M} = \frac{\sigma^2_{obs}-\sigma^2_0}{\sigma^2_0}.
\end{equation}

Obviously, an overestimate of the observed velocity dispersion immediately translates into an inflation of the dynamical mass of the system.\\
Given this, the main quantities we focus our attention on are the two expressions for the velocity dispersion $\sigma_{tot}$ (Eq. \ref{sigma_tot}) and the luminosity averaged $\sigma_{tot,lum}$ (Eq. \ref{sigma_tot_lum}), for they include the binary orbital motion, which becomes more and more important at increasing binary fraction and with the shrinking of the binary semi-major axis. 

\begin{figure}[h]
    \begin{subfigure}
        \centering
        \includegraphics[scale=0.3]{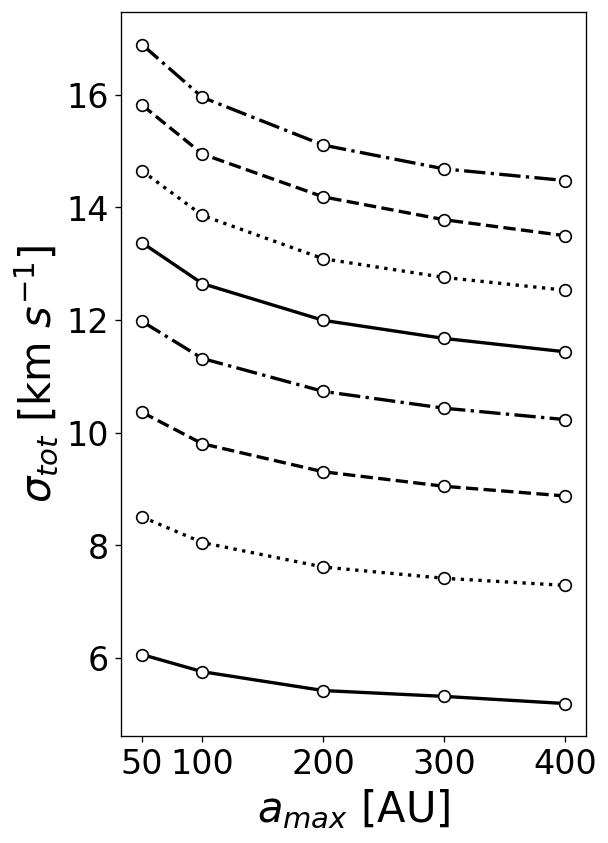}
    \end{subfigure}
    \hfill
    \begin{subfigure}
        \centering
        \includegraphics[scale=0.3]{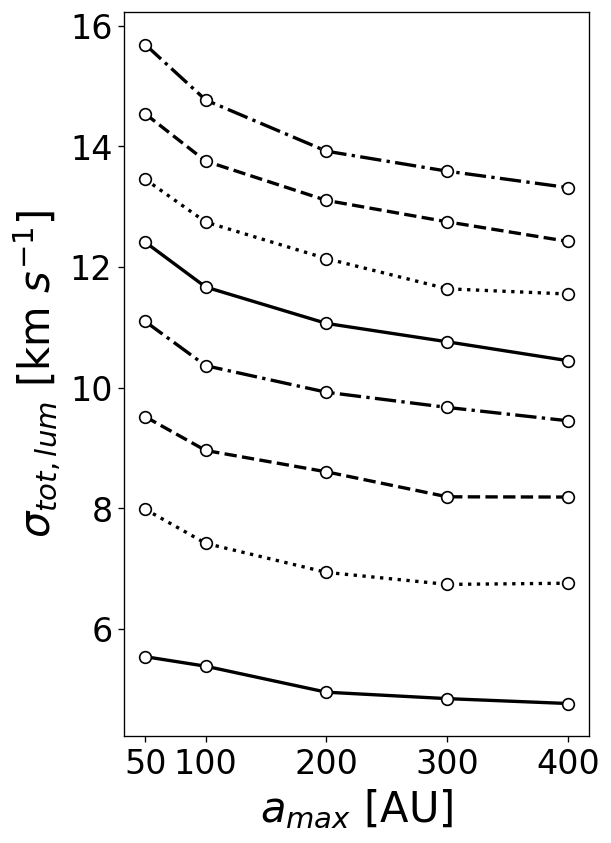}
    \end{subfigure}
    \newline
    \begin{subfigure}
        \centering
        \includegraphics[scale=0.3]{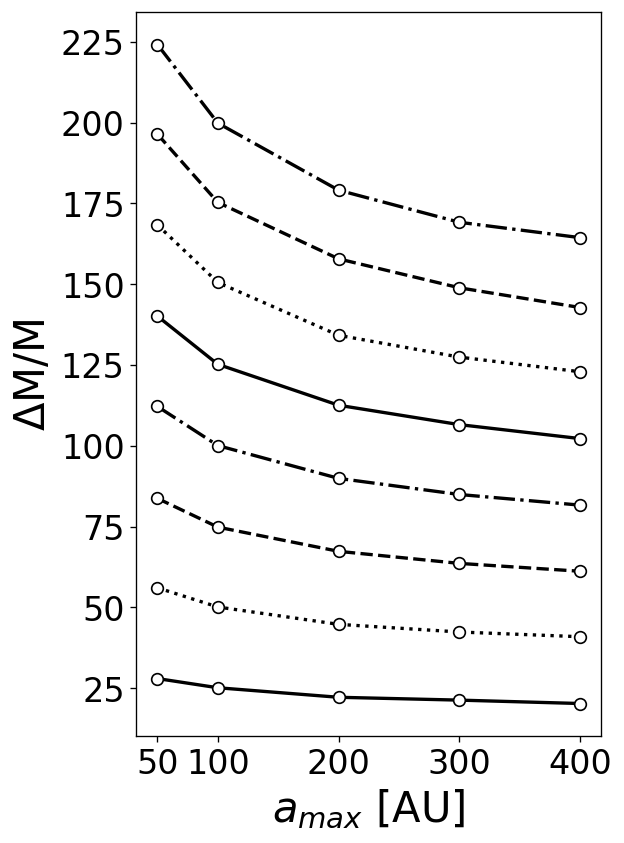}
    \end{subfigure}
    \hfill
    \begin{subfigure}
        \centering
        \includegraphics[scale=0.3]{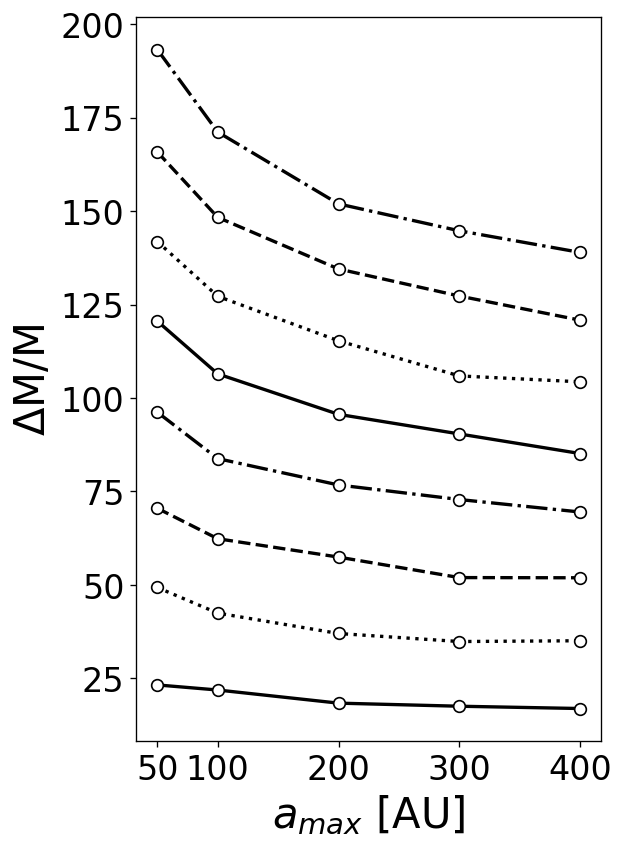}
    \end{subfigure}
    \caption{For the simulated UFD: upper panels illustrate the dependence of the velocity dispersion $\sigma_{tot}$ (top-left panel) and of the luminosity averaged velocity dispersion $\sigma_{tot,lum}$ (top-right panel) on the variation of the upper boundary $a_{max}$ of the binary semi-major axis distribution for the reference model.
    Lower panels show the corresponding relative mass difference $\Delta M/M$. Each curve corresponds to a different value of $f_b$ going bottom-up from 0.05 to 0.4 in steps of 0.05.}
    \label{UFD_a_max}
\end{figure}

\begin{figure}[h]
    \begin{subfigure}
        \centering
        \includegraphics[scale=0.3]{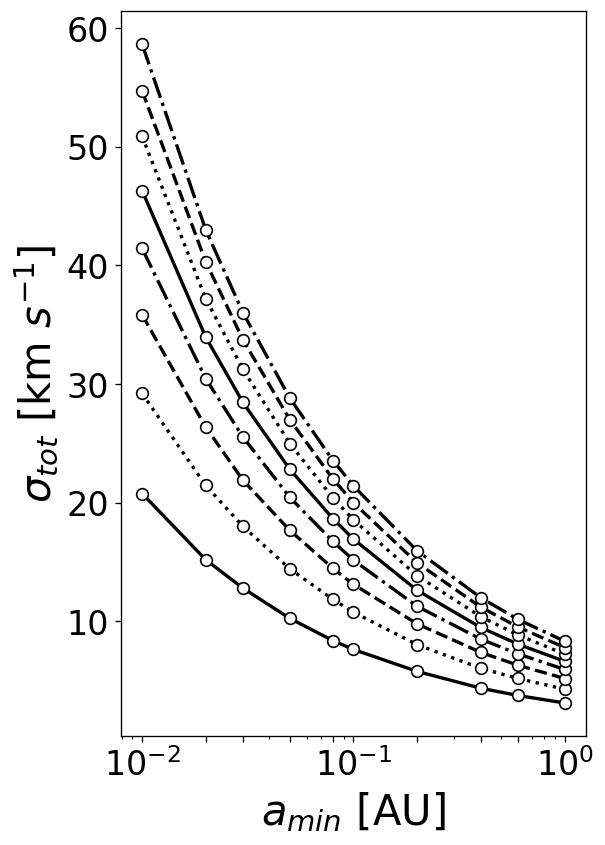}
    \end{subfigure}
    \hfill
    \begin{subfigure}
        \centering
        \includegraphics[scale=0.3]{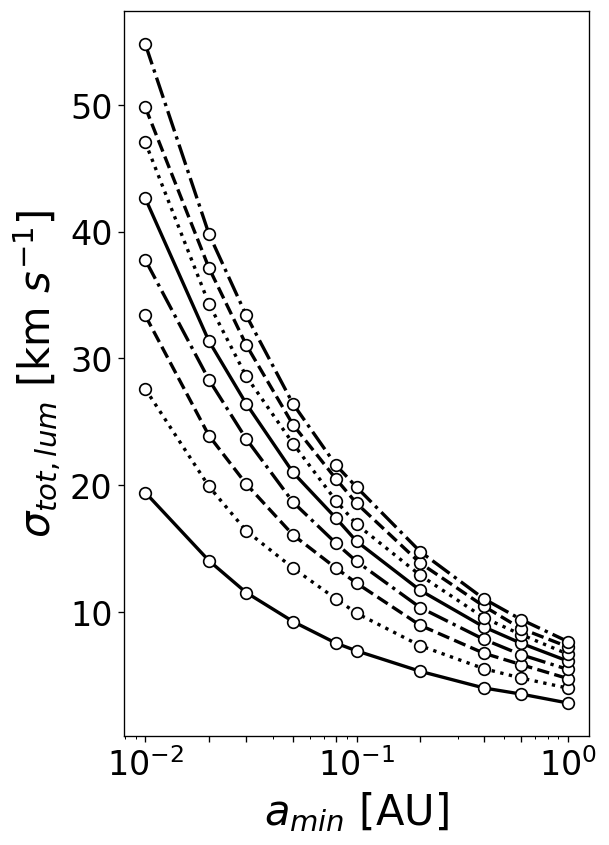}
    \end{subfigure}
    \newline
    \begin{subfigure}
        \centering
        \includegraphics[scale=0.3]{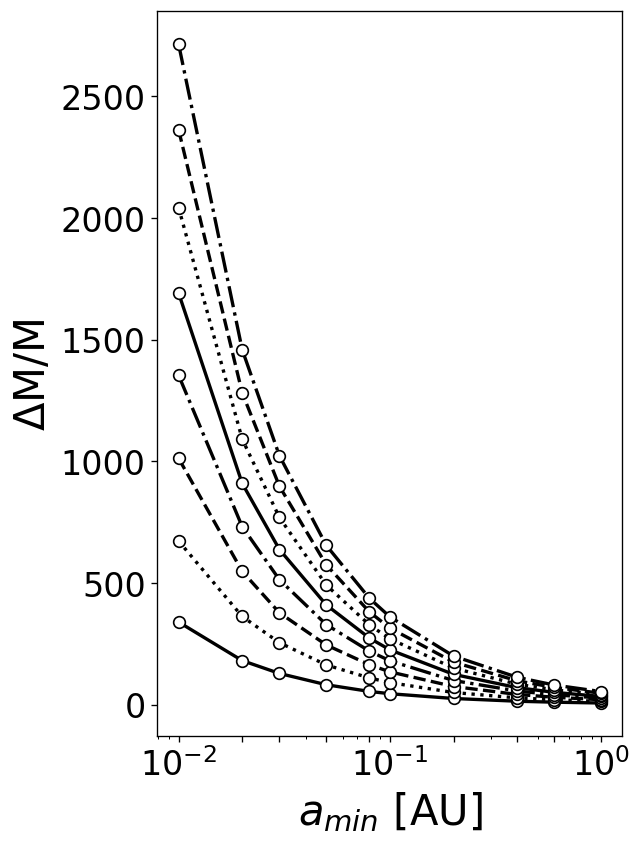}
    \end{subfigure}
    \hfill
    \begin{subfigure}
        \centering
        \includegraphics[scale=0.3]{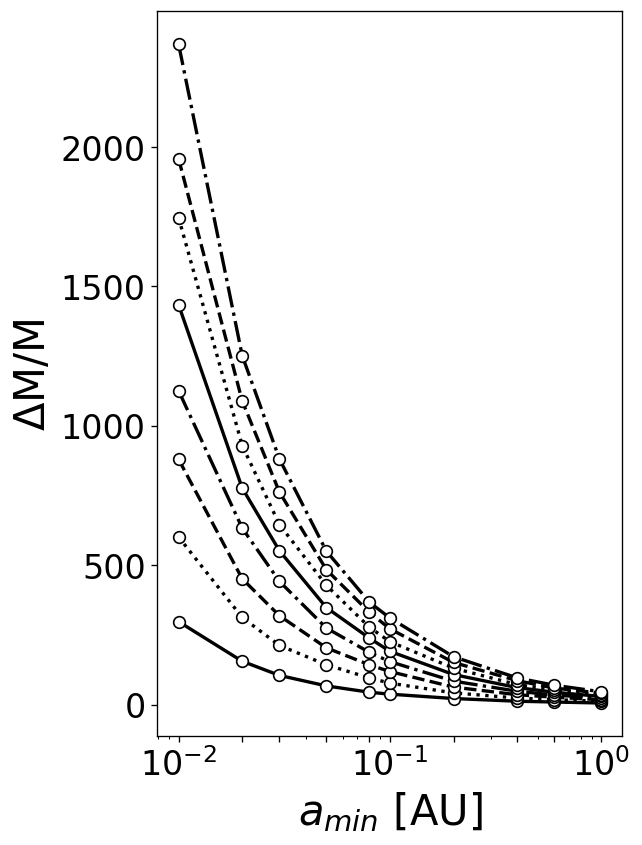}
    \end{subfigure}    
    \caption{As in Fig. \ref{UFD_a_max}, but for the variation of $a_{min}$ at fixed value of $a_{max} = 100$ AU.}
    \label{UFD_a_min}
\end{figure}

\begin{figure}[h]
    \begin{subfigure}
        \centering
        \includegraphics[scale=0.3]{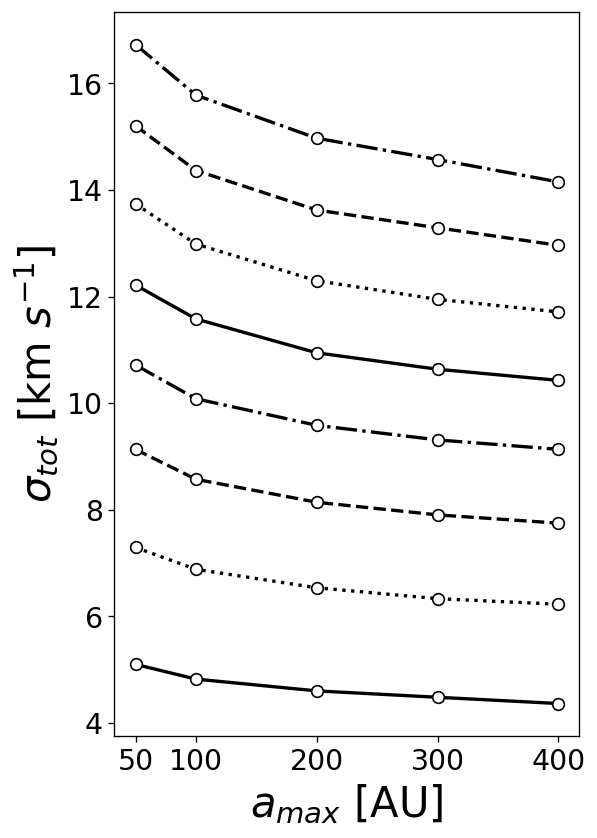}
    \end{subfigure}
    \hfill
    \begin{subfigure}
        \centering
        \includegraphics[scale=0.3]{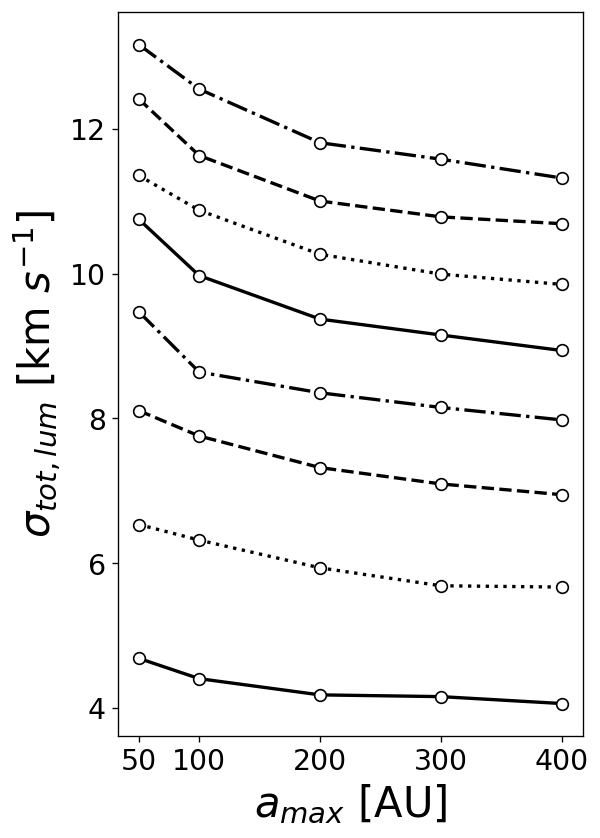}
    \end{subfigure}
    \newline
    \begin{subfigure}
        \centering
        \includegraphics[scale=0.3]{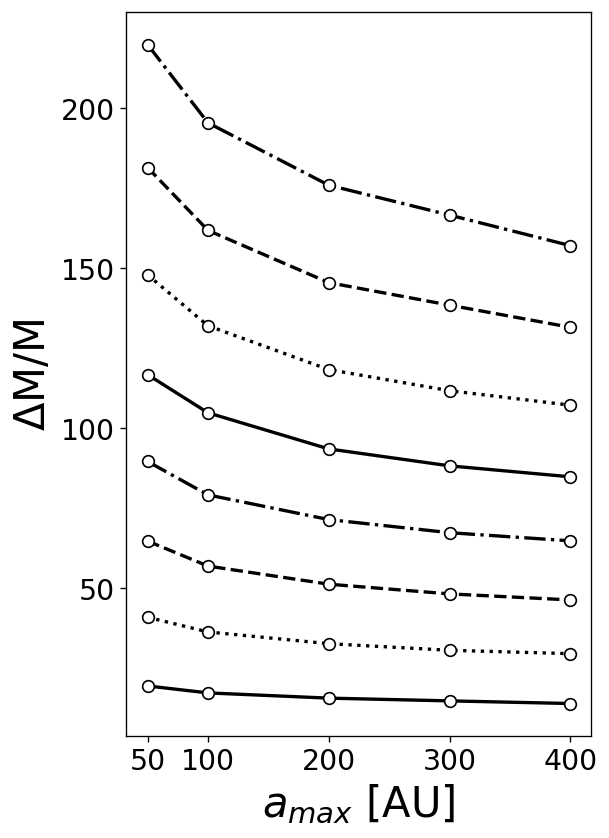}
    \end{subfigure}
    \hfill
    \begin{subfigure}
        \centering
        \includegraphics[scale=0.3]{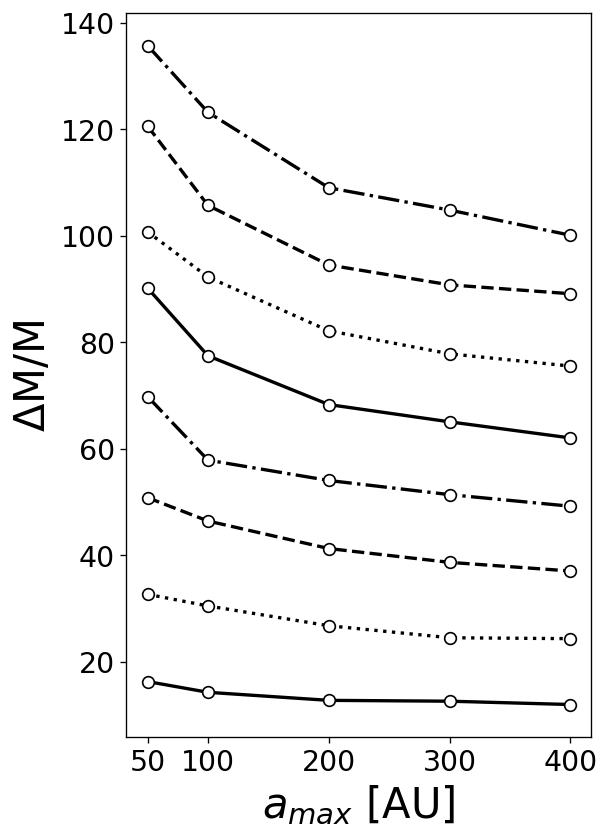}
    \end{subfigure}
    \caption{As in Fig. \ref{UFD_a_max}, but accounting for RLOF.}
    \label{UFD_a_max_RLOF}
\end{figure}

\begin{figure}[h]
    \begin{subfigure}
        \centering
        \includegraphics[scale=0.3]{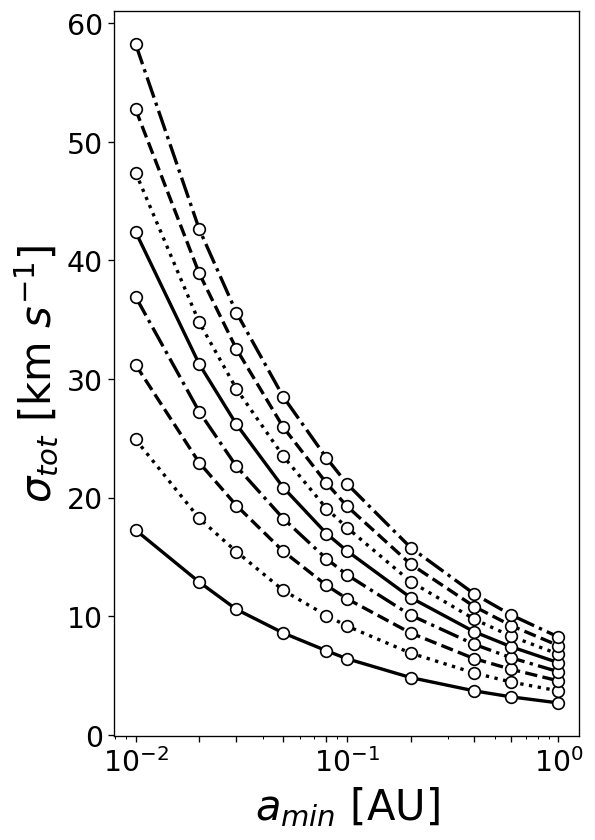}
    \end{subfigure}
    \hfill
    \begin{subfigure}
        \centering
        \includegraphics[scale=0.3]{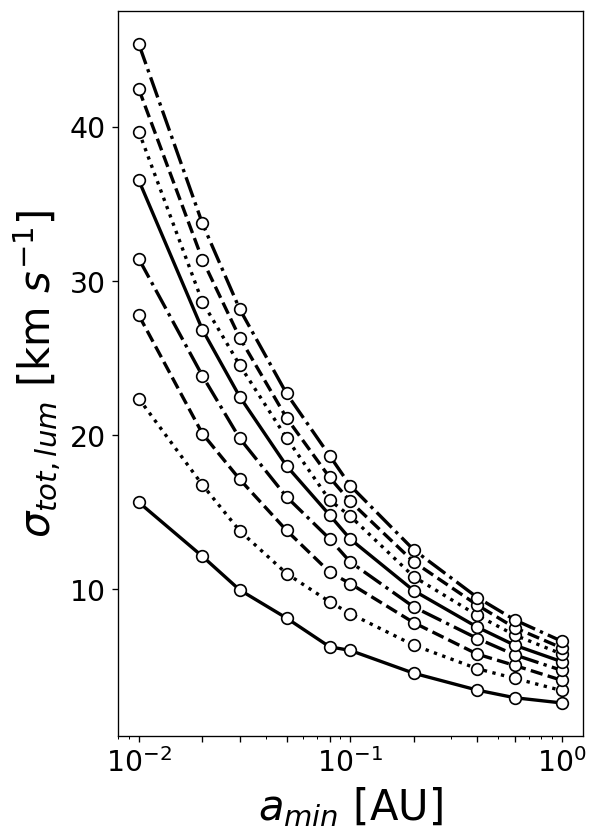}
    \end{subfigure}
    \newline
    \begin{subfigure}
        \centering
        \includegraphics[scale=0.3]{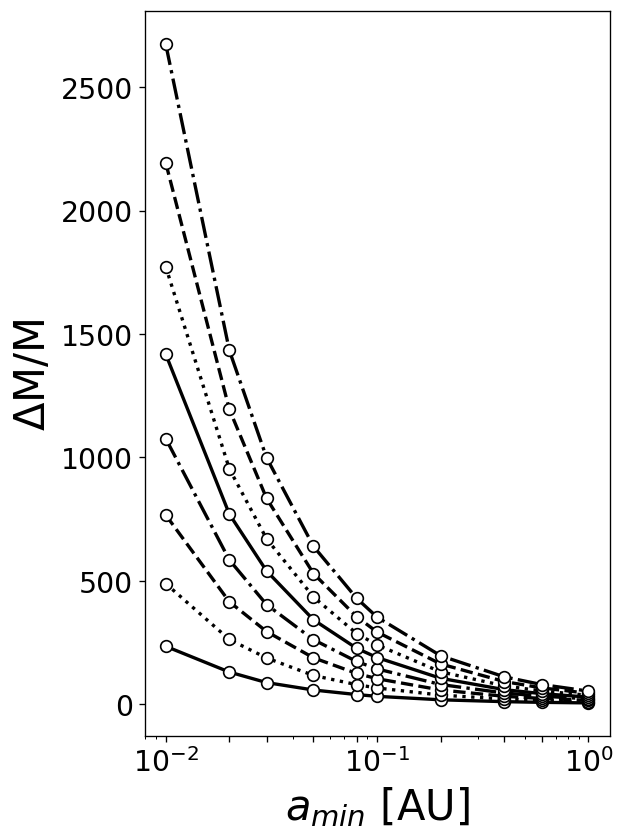}
    \end{subfigure}
    \hfill
    \begin{subfigure}
        \centering
        \includegraphics[scale=0.3]{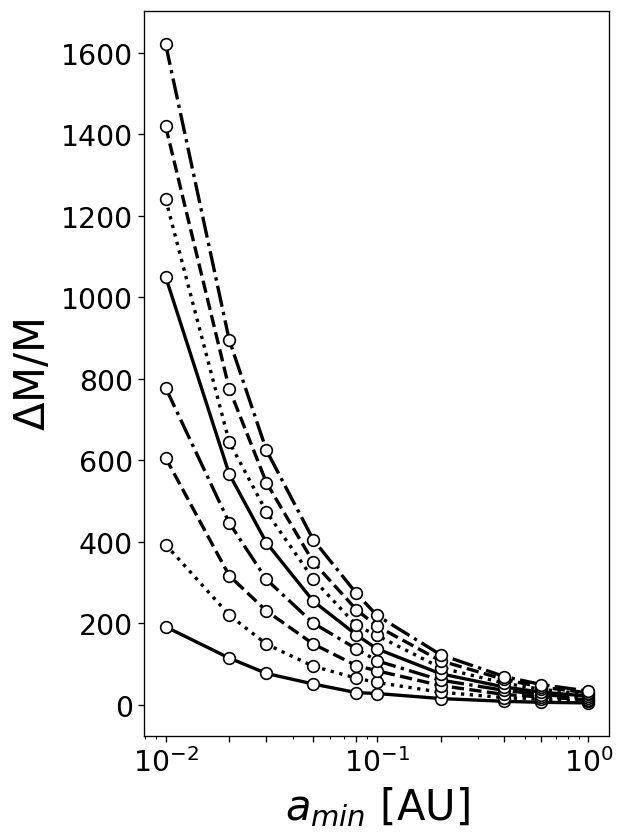}
    \end{subfigure}
    \caption{As in Fig. \ref{UFD_a_min}, but accounting for RLOF.}
    \label{UFD_a_min_RLOF}
\end{figure}

Fig. \ref{UFD_a_max} shows, for the simulated UFD, the role of the variation of $a_{max}$ in calculating $\sigma_{tot}$ (Fig. \ref{UFD_a_max}, top-left panel) and $\sigma_{tot,lum}$ (Fig. \ref{UFD_a_max}, top-right panel), and the related effect on the $\Delta M/M$ evaluation (Fig. \ref{UFD_a_max}, bottom-left panel, and Fig. \ref{UFD_a_max}, bottom-right panel).\\
Note that $\sigma_{tot,lum}$ is systematically smaller than $\sigma_{tot}$, thus yielding a corresponding lower estimate for the virial mass. Since the difference between $\sigma_{tot}$ and $\sigma_{tot,lum}$ reaches at most the $13\%$ for a binary fraction $f_b = 0.4$ in the case of our reference model ($a_{max} = 100$ AU), we deduce that the overall dependence of the observed velocity dispersion on $a_{max}$ is not very relevant.\\
On the contrary, we see from Fig. \ref{UFD_a_min} that the lessening of $a_{min}$ is much more important in inflating the velocity dispersion. In fact, when $a_{min} =$ 0.01 AU, $\sigma$ results larger than $20$ \kms even for $f_b=0.05$, and then increases approximately as $\sqrt{f_b}$. This implies a huge enhancement of the predicted virial mass as opposed to the real mass of the system, which is evident from the bottom panels of Fig. \ref{UFD_a_min}.\\
Fig. \ref{UFD_a_max} and Fig. \ref{UFD_a_min} must be compared, respectively, to Fig. \ref{UFD_a_max_RLOF} and Fig. \ref{UFD_a_min_RLOF}, which display the trend of $\sigma_{tot}$ and $\sigma_{tot,lum}$, as well as that of the associated $\Delta M/M$, when adding RLOF. 
As expected, we notice a modest, although global, decrease of the observed velocity dispersion; this is quite clear especially in the luminosity averaged case, where the velocity of merging binaries is weighted by the sum of their components' luminosities (see Eq. \ref{sigma_tot_lum_RLOF}). Still, if binaries are assumed to drop out of the sample when RLOF befalls the primary star only, regardless of whether an actual merger occurs (\citeauthor{1996AJ....111..750O}, \citeyear{1996AJ....111..750O}; \citeauthor{2010ApJ...721.1142M}, \citeyear{2010ApJ...721.1142M}), the observed velocity dispersion increases again to almost recover its original value, owing to the smaller number of rejected pairs.\\

\begin{figure}[h]
    \begin{subfigure}
        \centering
        \includegraphics[scale=0.3]{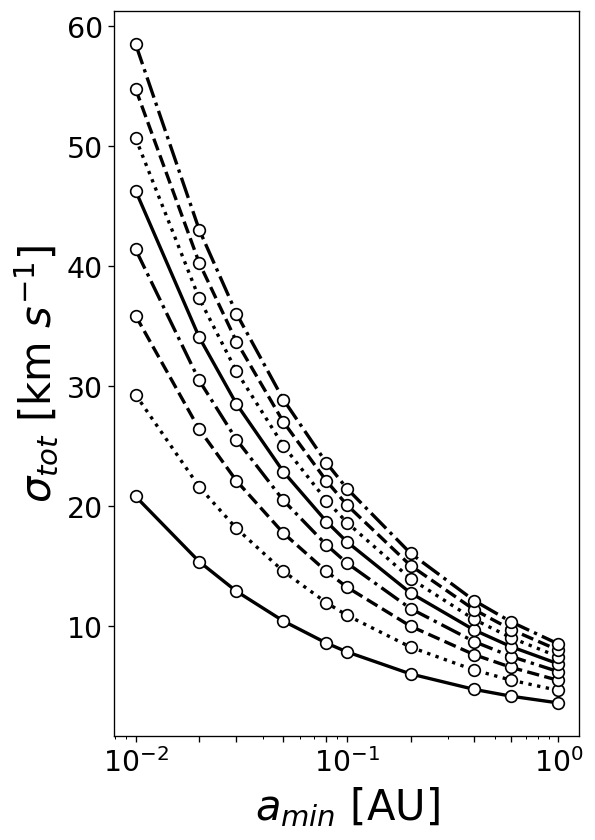}
    \end{subfigure}
    \hfill
    \begin{subfigure}
        \centering
        \includegraphics[scale=0.3]{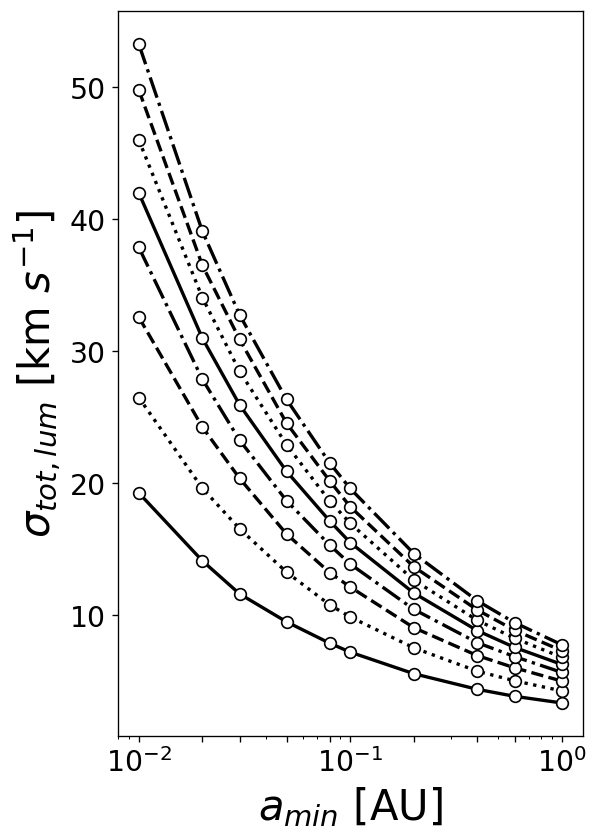}
    \end{subfigure}
    \newline
    \begin{subfigure}
        \centering
        \includegraphics[scale=0.3]{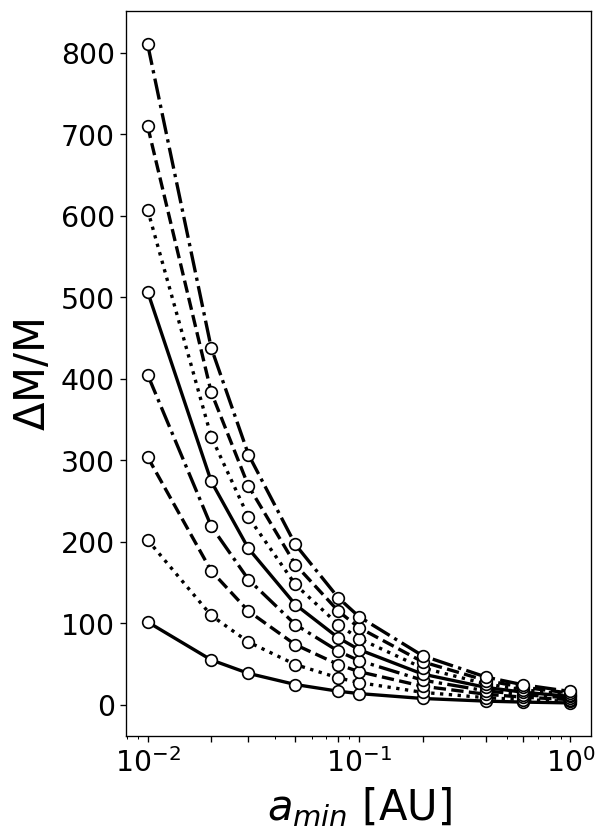}
    \end{subfigure}
    \hfill
    \begin{subfigure}
        \centering
        \includegraphics[scale=0.3]{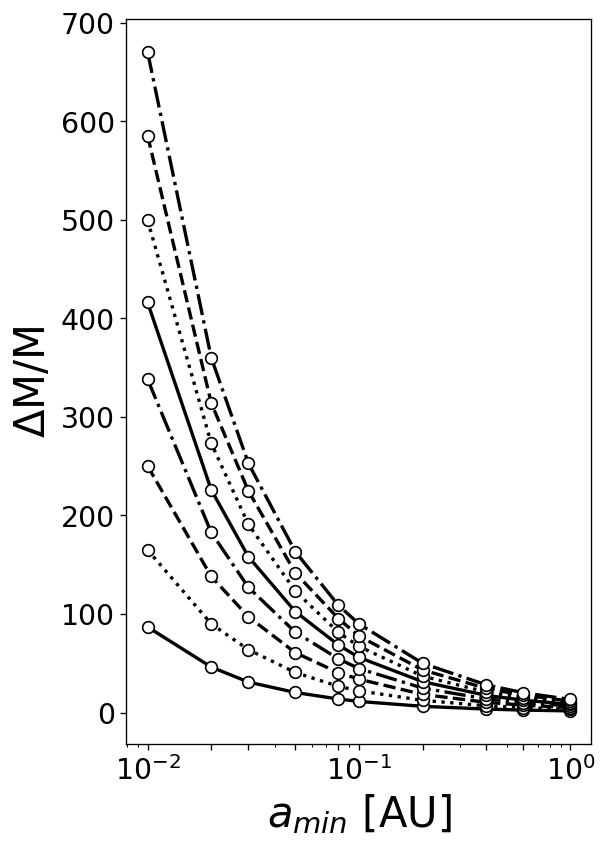}
    \end{subfigure}
    \caption{For the simulated dSph: upper panels illustrate the dependence of the velocity dispersion $\sigma_{tot}$ (top-left panel) and of the luminosity averaged velocity dispersion $\sigma_{tot,lum}$ (top-right panel) on the variation of the lower boundary $a_{min}$ of the binary semi-major axis distribution for the reference model.
    Lower panels show the corresponding relative mass difference $\Delta M/M$. Each curve corresponds to a different value of $f_b$ going bottom-up from 0.05 to 0.4 in steps of 0.05.}
    \label{dSph_a_min}
\end{figure}

\begin{figure}[h]
    \begin{subfigure}
        \centering
        \includegraphics[scale=0.3]{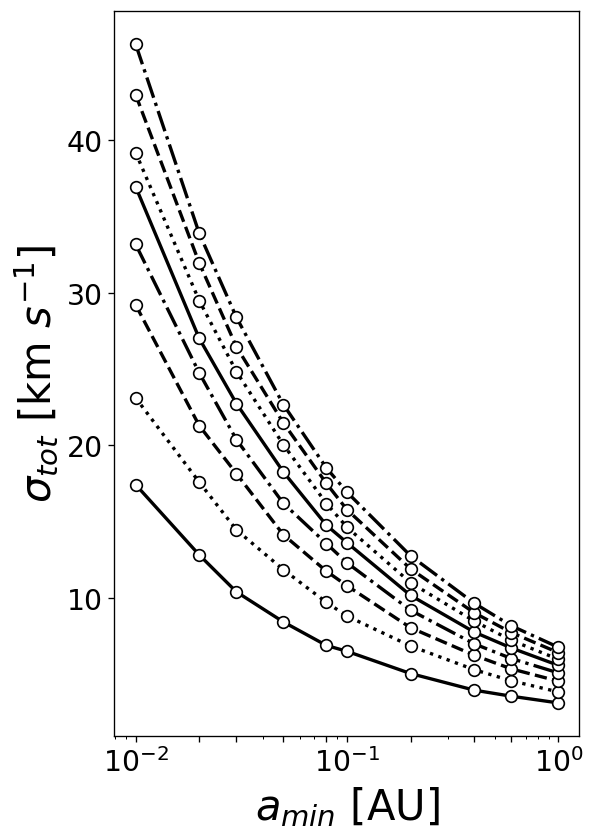}
    \end{subfigure}
    \hfill
    \begin{subfigure}
        \centering
        \includegraphics[scale=0.3]{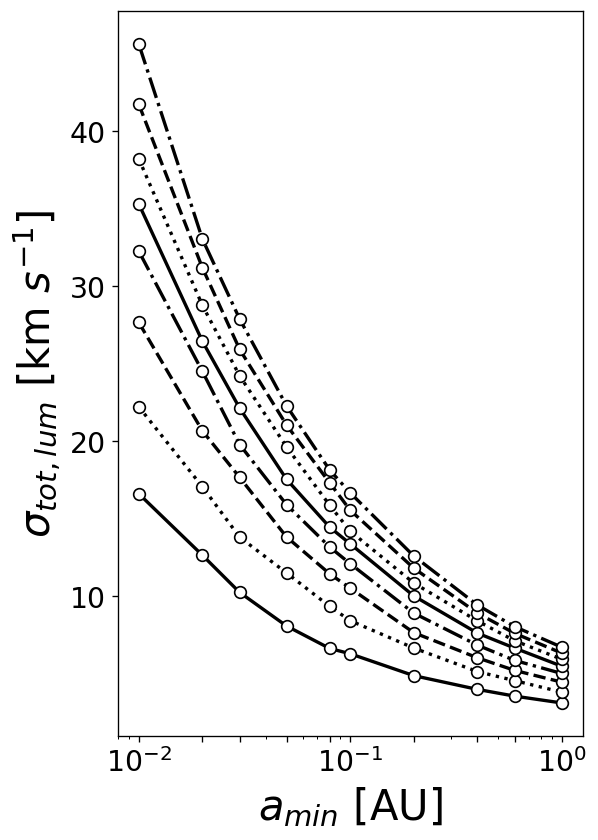}
    \end{subfigure}
    \newline
    \begin{subfigure}
        \centering
        \includegraphics[scale=0.3]{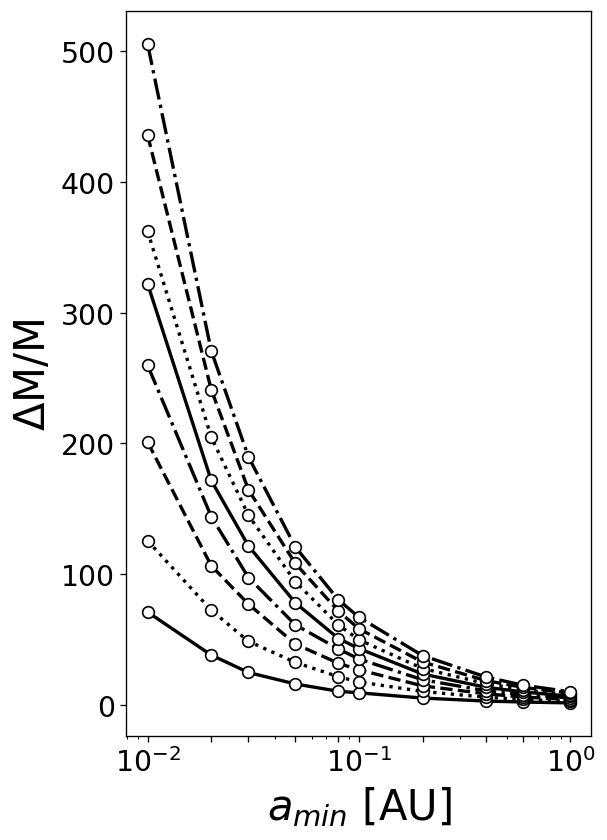}
    \end{subfigure}
    \hfill
    \begin{subfigure}
        \centering
        \includegraphics[scale=0.3]{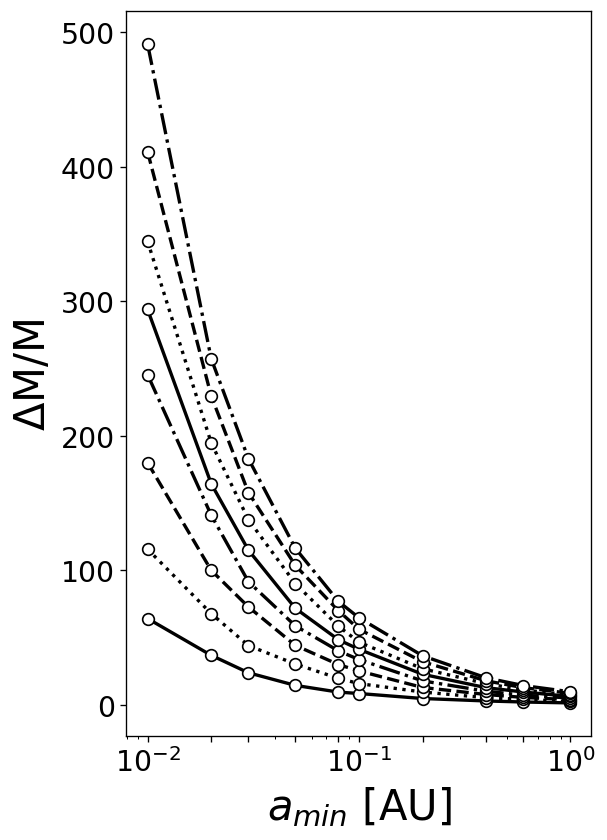}
    \end{subfigure}
    \caption{As Fig. \ref{dSph_a_min}, but accounting for both RLOF and the luminosity cut-off.}
    \label{dSph_a_min_RLOF_lum_cut_off}
\end{figure}

With regards to our model dSph, we report only the most meaningful results in Fig. \ref{dSph_a_min} and Fig. \ref{dSph_a_min_RLOF_lum_cut_off}, which show, respectively, the dependence of $\sigma$ and the related $\Delta M/M$ on the variation of $a_{min}$ before and after imposing the aforementioned cuts. A straightforward comparison of Fig. \ref{dSph_a_min} with Fig. \ref{UFD_a_min}, and of Fig. \ref{dSph_a_min_RLOF_lum_cut_off} with Fig. \ref{UFD_a_min_RLOF} corroborates our expectation that the boost of the global velocity dispersion caused by binaries is more prominent in bigger systems (like dSphs) than in UFDs. Note, \textit{inter alia}, that the binary fraction slightly increases due to the luminosity cut-off, since it affects single stars more than binaries. In reference to Fig. \ref{dSph_a_min_RLOF_lum_cut_off}, the new binary fraction, i.e., $f_b \in [0.09,0.16,0.23,0.29,0.33,0.37,0.41,0.44]$, is indeed higher with respect to the original case.\\
In addition to this, we point out that, in line with the predictions by \citeauthor{2020ApJ...896..152R} (\citeyear{2020ApJ...896..152R}), although in the different context of open star clusters (OCs), the luminosity cut-off is not much impactful on the velocity dispersion estimate. In fact, we found that the observed velocity dispersion experiences the most dramatic decline as a consequence of the RLOF rejection, not the luminosity cut-off, which provokes a further reduction of $\sim 1-5$ \kms~ with increased binary fraction. In particular, as for our reference model ($a_{min}=0.2$ AU), the lessening of $\sigma_{tot}$ goes from $\sim 25\%$ ($f_b = 0.4$) to $\sim 40\%$ ($f_b = 0.1$), whereas that of $\sigma_{tot,lum}$ from $\sim 20\%$ to $\sim 35\%$ for the same values of $f_b$.

Finally, we examined the role of mass coupling in binaries through a comparison between the outcomes relative to the random pairing procedure and those coming from the assumption of a power-law mass ratio distribution $p(q)\propto q^{-0.4}$. Being a complete compatibility of a given mass function with a given binary mass ratio distribution impossible, we decided to implement the power-law mass distribution by normalizing the $(m_1,m_2)$ mass pairs to give the same binary total mass $m_1+m_2$ of the random pairing case.
At fixed $a_{max}$ of the standard model and varying $a_{min}$ in the usual range, the values of $\sigma^2$ computed in the case of power-law mass ratio distribution differ from the ones of random pairing for $\sim \pm 20 \%$. A similar variation range is found if $a_{min}$ is kept fixed and $a_{max}$ is varied.
As a net result, the choice of a power-law distribution leads to an average underestimate of $\sigma^2$ of the order of $15\%$: this has little effect on the conclusions of our work, which are in the direction of pointing out the importance of binaries in the dynamical mass estimate of a stellar system like those studied here.

\subsection{Dependence of the results on the system mass and scale radius}

\begin{figure}[h]
    \centering
    \begin{subfigure}
        \centering
        \includegraphics[scale=0.35]{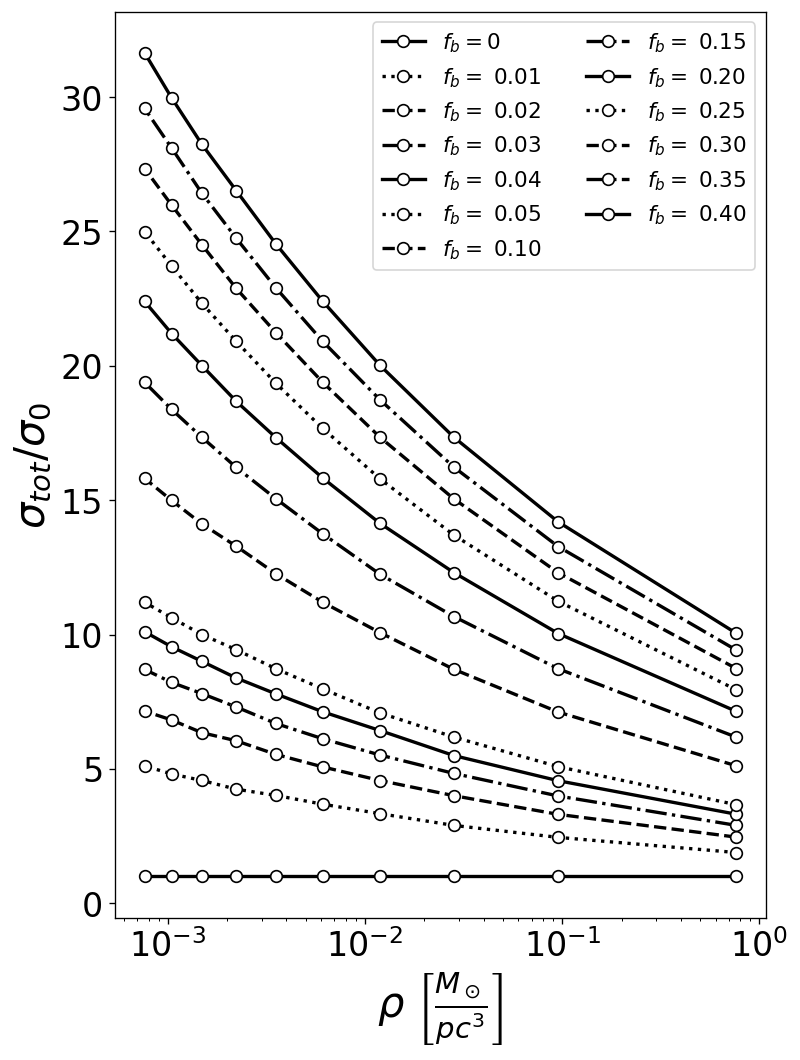}
    \end{subfigure}
    \hfill
    \begin{subfigure}
        \centering
        \includegraphics[scale=0.35]{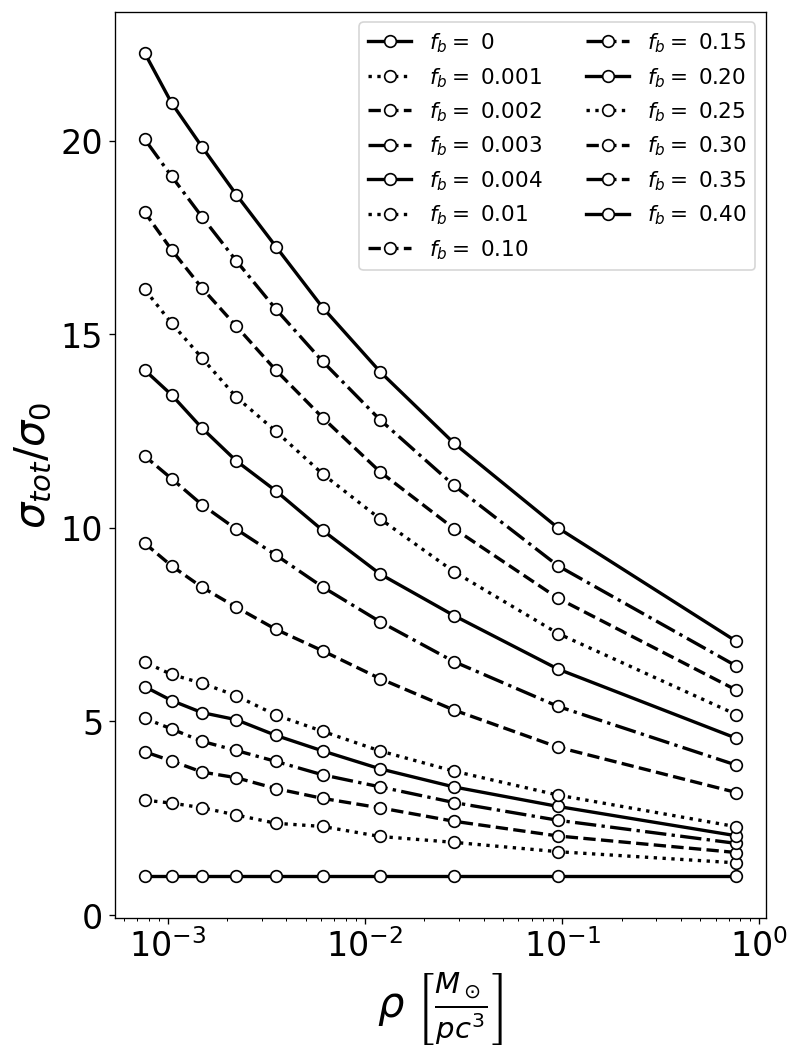}
    \end{subfigure}
    \hfill
    \caption{Dependence of the ratio $\sigma_{tot}/\sigma_0$ on the mean mass density of the system obtained by varying the scale radius $R$ from 25 to 250 pc with steps of 25 at fixed total mass $M=5 \times 10^4$ M$_\odot$. The values of $f_b$ label each curve going bottom-up, according to the legend, and differ depending on whether RLOF is taken into account (lower panel) or not (upper panel) in the calculation of $\sigma_{tot}$.}
    \label{sigma_ratio_density}
\end{figure}

In order to perform a more comprehensive investigation of the impact of the binary content in small size, low dense stellar systems, we placed binary stars in ever-decreasing density dwarf galaxies by extending the scale radius from 25 to 250 pc, with steps of 25 pc, for the fixed total mass $M = 5 \times 10^4$ M$_\odot$ we assumed to represent a UFD in our simulations.\\
Fig. \ref{sigma_ratio_density} (upper panel) shows the decreasing trend of $\sigma_{tot}/\sigma_0$ as a function of the mean mass density of the system without taking account of RLOF.
Interestingly, in the case of our reference model ($\rho \simeq 0.1$ M$_\odot$ pc$^{-3}$) a binary fraction of just $5\%$ suffices to produce a significant enhancement of the dynamical mass (of a factor of $\sim 25.8$ for the above-mentioned instance).\\
Furthermore, we emphasize that we essentially recover the results of \citeauthor{2010ApJ...721.1142M} (\citeyear{2010ApJ...721.1142M}), who predicted that, in dSphs with $\sigma_{obs} > 4$ \kms, the inflation due to binary orbital motion would unlikely exceed the 30\%. Now, we obtain $\sigma_{obs} = 4$ \kms for a binary fraction $f_b = 0.03$ and, since the intrinsic velocity dispersion $\sigma_0$ goes from $\sim 1.6$ to $\sim 0.5$ \kms at increasing scale radius of the system, it follows that the overestimate of the observed velocity dispersion reaches at most the 8\%. Inversely, in the case of higher binary fractions, for which  $\sigma_{obs}$ is larger than 4 \kms, such an inflation grows exactly up to $\sim 30\%$.
These considerations hold if RLOF is accounted for (Fig. \ref{sigma_ratio_density}, lower panel), because $\sigma_{tot}$ decreases of less than $\sim 1\%$ with respect to the corresponding values in the absence of RLOF.\\
Note that, being $f_b \leq 0.4$ in our analysis, we can obviously argue that the threshold suggested by \citeauthor{2010ApJ...721.1142M} (\citeyear{2010ApJ...721.1142M}) for the boost of the observed velocity dispersion may be overtaken if a more numerous binary population with our characteristics is considered. Yet, we exercise particular caution in this respect, as aware of the differences in our modeling approach, especially regarding the choice of the binary velocity and period distributions.\\
Moreover, the dependence of $\sigma_{tot}/\sigma_0$ on the mean mass density explains why in systems like GCs, which are small sized but also dense, there is no expectation for a relevant $\sigma$ inflation due to binaries. Incidentally, GCs are deemed to be totally deprived of DM.\\
Note in addition that, contrary to dwarf galaxies, where the intrinsic non-collisionality would lead to an almost constant in time $f_b$, GCs are presently supposed to contain only a limited fraction of binaries owing to their collisional nature (\citeauthor{Milone+2012}, \citeyear{Milone+2012}). Thereby, being the destruction rate of binary stars through dynamical interactions higher than the formation one (\citeauthor{1992PASP..104..981H}, \citeyear{1992PASP..104..981H}),
we can state that detecting a significant enhancement of the observed velocity dispersion in these environments is very unlikely.\\
Our conclusion is enforced by the calculation of the half-mass relaxation time according to Eq. \ref{t_rh} (see Tab. \ref{structural_parameters}). 

\subsection{Mass-to-light ratio}

As we said in Sect. 3.1, for a given set of binary characteristics the dynamical mass estimation, $M_{dyn}$, is a linear function of $f_b$: 

\begin{equation}
    M_{dyn} = A + B f_b.
\end{equation}

In particular, for the simulated UFD, the values of the coefficients are $A = 5.55 \times 10^4$ M$_\odot$ and $B = 2.50 \times 10^7$ M$_\odot$ in the original set-up, whereas $A = -7.41 \times 10^5$ M$_\odot$ and $B = 2.53 \times 10^7$ M$_\odot$ when considering RLOF. For the simulated dSph, these coefficients are $A = 9.90 \times 10^6$ M$_\odot$ and $B = 1.50 \times 10^9$ M$_\odot$ when RLOF and the luminosity cut-off are not taken into account, while $A = -3.89 \times 10^7$ M$_\odot$ and $B = 9 \times 10^8$ M$_\odot$ when both of them are considered.\\
We then computed both the mass-to-bolometric light ratio, and the mass-to-light ratio in the V and B band for selected binary fractions in the case of our reference model (see Tab. \ref{mass-to-light_ratio} and Tab. \ref{mass-to-light_ratio_cut}).
Most notably, with regards to the B and V bands, it emerges (see Tab. \ref{mass-to-light_ratio}) that, for small sized systems such as UFDs, high values of $M/L$ arise even in the presence of a modest binary population, over-passing $100$ for $f_b >0.3$.
Of course, performing the RLOF rejection causes $M/L$ to diminish, being the total luminosity fixed.\\
Our findings are validated by a comparison with Fig. 4 (right panel) of \citeauthor{2019ARA&A..57..375S} (\citeyear{2019ARA&A..57..375S}), which displays the trend of the mass-to-light ratio within the half-light radius for a sample of UFDs as a function of the luminosity in the V band. Here we notice that, for a luminosity $L \sim 10^4$ L$_{V,\odot}$, i.e., the one associated to our simulated UFD, $(M/L)_V$ ranges from $\sim 10^2$ to $\sim 10^4$ M$_\odot$/L$_{V,\odot}$, in accordance with the predictions of our reference model for $f_b \geq 0.3$; this is true also in the event of RLOF, since the mass-to-light ratio is slightly reduced. Nevertheless, we stress, for the sake of clarity, that the mass-to-light ratio estimates associated to the UFDs for which velocity dispersion measurements are available, are affected by large uncertainties in the aforementioned luminosity regime, and that the dynamical mass has been calculated by following the prescription of \citeauthor{2010MNRAS.406.1220W} (\citeyear{2010MNRAS.406.1220W}), which may be a possible source of discrepancy with our results.\\
In closing, we put into evidence that, after the application of the cut procedure, the value of $\sigma_{tot}$ for the actual binary fraction $f_b = 0.37$ in our reference model dSph is magnified by a factor of $\sim 5.5$ with respect to the intrinsic value $\sigma_0 \simeq 2$ \kms;
this is consistent with the observations made by \citeauthor{2018AJ....156..257S} (\citeyear{2018AJ....156..257S}), who predicted a non-negligible effect of Leo II-like binary fractions in galaxies having $\sigma_0 \simeq 0.5-2$ \kms. Even so, as highlighted by \citeauthor{2016MNRAS.463.1865D} (\citeyear{2016MNRAS.463.1865D}), such an influence is tightly related to the total luminosity of the system, provided that virial equilibrium is assumed, and becomes much more pronounced when $L \leq 10^6$ L$_{V,\odot}$. 
This is a natural outcome of a velocity dispersion inflation as due to a given binary population, which is, of course, fractionally more important in lighter systems than in larger. Therefore, according to \citeauthor{2016MNRAS.463.1865D} (\citeyear{2016MNRAS.463.1865D}), we expect that, for the considered dSph total luminosity $L \simeq 10^7$ L$_{V,\odot}$, binaries alone would not be able to boost the observed velocity dispersion to the extent that the presence of DM may be totally ruled out. Indeed, $(M/L)_V$ corresponding to $f_b = 0.33$ for the simulated dSph (see Tab. \ref{mass-to-light_ratio_cut}) undergoes a minor enhancement owing to the sole action of binaries, if compared to the UFD case, where the total luminosity is set at $L \simeq 10^4$ L$_{V,\odot}$. We consequently find confirmation that binary stars affect the internal dynamics of UFD to a greater degree than dSph, which may be unlikely regarded as utterly composed of baryonic matter.

\begin{deluxetable}{c c c c c}[h]
\tablecaption{Values of the mass-to-light ratio in the bolometric and $V$ and $B$ photometric bands for various binary fractions, in the case of our reference model.} 
\label{mass-to-light_ratio}
\tablehead{
\colhead{Object} & \colhead{$f_b$} & \colhead{$(M_{dyn}/L)_{bol}$} & \colhead{$(M_{dyn}/L)_V$} & \colhead{$(M_{dyn}/L)_B$} \\
\colhead{} & \colhead{} & \colhead{(M$_\odot$/L$_\odot$)} & \colhead{(M$_\odot$/L$_{V,\odot}$)} & \colhead{(M$_\odot$/L$_{B,\odot}$)}
}
\startdata
dSph & 0 & 0.07 & 0.73 & 0.60 \\
 & 0.05 & 0.63 & 6.15 & 5.06 \\
 & 0.15 & 1.75 & 17.07 & 14.03 \\
 & 0.30 & 3.41 & 33.38 & 27.44 \\
 & 0.40 & 4.53 & 44.29 & 36.40 \\
\hline
UFD & 0 & 0.07 & 0.73 & 0.60 \\
 & 0.05 & 1.95 & 19.04 & 15.65 \\
 & 0.15 & 5.64 & 55.16 & 45.34 \\
 & 0.30 & 11.26 & 110.13 & 90.53 \\
 & 0.40 & 14.92 & 145.88 & 119.95 \\
\enddata
\end{deluxetable}

\begin{deluxetable}{c c c c c}[h]
\tablecaption{As Tab. \ref{mass-to-light_ratio}, but accounting for
both RLOF and the luminosity cut-off (dSph case) and RLOF only for  the UFD case. For the dSph, $f_b$ refers to the actual binary fraction obtained after the luminosity cut procedure.}
\label{mass-to-light_ratio_cut}
\tablehead{
\colhead{Object} & \colhead{$f_b$} & \colhead{$(M_{dyn}/L)_{bol}$} & \colhead{$(M_{dyn}/L)_V$} & \colhead{$(M_{dyn}/L)_B$} \\
\colhead{} & \colhead{} & \colhead{(M$_\odot$/L$_\odot$)} & \colhead{(M$_\odot$/L$_{V,\odot}$)} & \colhead{(M$_\odot$/L$_{B,\odot}$)}
}
\startdata
dSph & 0 & 0.07 & 0.73 & 0.60 \\
 & 0.09 & 0.44 & 4.33 & 3.56 \\
 & 0.23 & 1.13 & 11.07 & 9.10 \\
 & 0.37 & 2.19 & 21.46 & 17.64 \\
 & 0.44 & 2.68 & 27.27 & 22.41 \\ 
\hline
UFD & 0 & 0.07 & 0.73 & 0.60 \\
 & 0.05 & 1.36 & 13.34 & 10.96 \\
 & 0.15 & 4.31 & 42.12 & 34.63 \\
 & 0.30 & 9.88 & 96.65 & 79.45 \\
 & 0.40 & 14.69 & 142.63 & 117.27 \\
\enddata
\end{deluxetable}

\section{Conclusions} \label{conc}

We studied the role of unresolved binary stars in inflating the observed velocity dispersion of dwarf galaxies by realizing a set of non-dynamical simulations in dependence on various binary system parameters.\\
At odds with previous investigations where sophisticated statistical analyses were performed (\citeauthor{2010ApJ...721.1142M}, \citeyear{2010ApJ...721.1142M}; \citeauthor{2018AJ....156..257S}, \citeyear{2018AJ....156..257S}), in this first application of our model we took into account the explicit influence of each orbital element, and explored conservative regions of the parameter space.
We considered two different spherical systems aiming at representing a typical dwarf spheroidal galaxy (dSph) and an ultra-faint dwarf (UFD) galaxy. We drew our attention to the effects of the variation of binary orbital parameters, obtaining, as principal result, that the dominant impact on the estimate of the system velocity dispersion, in the hypothesis of an unresolved binary population, is given by the semi-major axis (and so by the orbital period) distribution.

The main outcomes of this study can be summarized as follows:
\begin{itemize}
    \item the presence of an abundant quantity of unresolved binaries with relatively low periods (see Tab. \ref{simulations_summary}) leads to a significant enhancement of the observed velocity dispersion, and, consequently, of the dynamical mass evaluated through the virial theorem upon assumption of stationary systems. This result differs from  \citeauthor{2019MNRAS.487.2961M} (\citeyear{2019MNRAS.487.2961M}), who assert, referring to the galaxy Reticulum II, that a high fraction of close binaries in low-metallicity environments, such as UFDs, is unable to make an appreciable contribution to the observed velocity dispersion;
    \item the observed squared velocity dispersion is a linear function of the binary fraction, as outlined, e.g., by \citeauthor{2010ApJ...721.1142M} (\citeyear{2010ApJ...721.1142M});
    \item the corresponding mass estimate is inflated with respect to the real mass of the system, and increases with both the binary fraction and the shrinking of the binary semi-major axis (i.e., by diminishing the binary orbital periods);
    \item low-mass systems (UFDs) suffer more from the contribution of a given binary population due to their smaller intrinsic velocity dispersion ($\sigma \propto \sqrt{M}$);
    \item the action of RLOF translates into a modest reduction of the observed velocity dispersion in both our simulated galaxies. However, its decline is more prominent in the model dSph, given the additional luminosity cut-off, which involves single and binary stars differently;
    \item the introduction of a power-law mass ratio distribution $p(q)\propto q^{-0.4}$ for the binary mass coupling causes $\sigma^2$ to be underestimated of $\sim 15\%$ with respect to the random pairing case, hence affecting in a modest way the evaluation of the dynamical mass; 
     \item the boost of the observed velocity dispersion by binary stars is a steeply decreasing function of the mean mass density of the system. In particular, for low-density galactic hosts, even a small fraction ($5\%$) of binaries with our standard characteristics produces a non-negligible inflation of the dynamical mass (i.e., by a factor of $\sim 25.8$ in the case of our reference model without accounting for RLOF);
    \item the values of the mass-to-light ratio we obtained are large and look compatible with those estimated observationally for UFDs and dSphs, offering, in the case of UFDs, an interpretation based on unresolved binaries as alternative or, at least, complementary to that of an overabundance of non-baryonic dark matter in such low density systems. 
\end{itemize}

In conclusion, our model provides a realistic and physically consistent explanation of the role of binary stars in the dynamical mass estimate of stellar systems, with the ultimate purpose of challenging the claim that only the presence of vast amounts of dark matter is of primary importance in this context.\\
We are aware that more robust and precise results require several improvements in both theoretical modelization and spectroscopic data availability, especially related to UFDs. Thus, while waiting for future observational facilities, we reserve to upgrade our model by accounting for the effects not only of stellar evolution (i.e., mass loss) and dynamics, but also of close interactions between binary components, in order to give a full-time picture of our mocked galaxies.\\
All these issues will be covered in a follow-up of this work.

\begin{acknowledgments}
C. Pianta and G. Carraro have been supported in this work by Padova University grant BIRD191235/19: {\it Internal dynamics of Galactic star clusters in the Gaia era: binaries, blue stragglers, and their effect in estimating dynamical masses}. The authors express their gratitude to the anonymous reviewer of this work for his/her useful comments and suggestions.
\end{acknowledgments}

\bibliography{bibliography}{}

\begin{thebibliography}{}
\expandafter\ifx\csname natexlab\endcsname\relax\def\natexlab#1{#1}\fi

\bibitem[{{Aaronson} \& {Olszewski}(1988)}]{1988IAUS..130..409A}
{Aaronson}, M., \& {Olszewski}, E.~W. 1988, in Large Scale Structures of the
  Universe, ed. J.~{Audouze}, M.~C. {Pelletan}, A.~{Szalay}, Y.~B.
  {Zel'dovich}, \& P.~J.~E. {Peebles}, Vol. 130, 409

\bibitem[{{Aarseth} {et~al.}(1974){Aarseth}, {Henon}, \&
  {Wielen}}]{1974A&A....37..183A}
{Aarseth}, S.~J., {Henon}, M., \& {Wielen}, R. 1974, \aap, 37, 183

\bibitem[{{Amorisco} \& {Evans}(2011)}]{2011MNRAS.411.2118A}
{Amorisco}, N.~C., \& {Evans}, N.~W. 2011, \mnras, 411, 2118

\bibitem[{{Belokurov} {et~al.}(2007){Belokurov}, {Zucker}, {Evans}, {Kleyna},
  {Koposov}, {Hodgkin}, {Irwin}, {Gilmore}, {Wilkinson}, {Fellhauer},
  {Bramich}, {Hewett}, {Vidrih}, {De Jong}, {Smith}, {Rix}, {Bell}, {Wyse},
  {Newberg}, {Mayeur}, {Yanny}, {Rockosi}, {Gnedin}, {Schneider}, {Beers},
  {Barentine}, {Brewington}, {Brinkmann}, {Harvanek}, {Kleinman}, {Krzesinski},
  {Long}, {Nitta}, \& {Snedden}}]{2007ApJ...654..897B}
{Belokurov}, V., {Zucker}, D.~B., {Evans}, N.~W., {et~al.} 2007, \apj, 654, 897

\bibitem[{{Belokurov} {et~al.}(2009){Belokurov}, {Walker}, {Evans}, {Gilmore},
  {Irwin}, {Mateo}, {Mayer}, {Olszewski}, {Bechtold}, \&
  {Pickering}}]{2009MNRAS.397.1748B}
{Belokurov}, V., {Walker}, M.~G., {Evans}, N.~W., {et~al.} 2009, \mnras, 397,
  1748

\bibitem[{{Dabringhausen} {et~al.}(2016){Dabringhausen}, {Kroupa}, {Famaey}, \&
  {Fellhauer}}]{2016MNRAS.463.1865D}
{Dabringhausen}, J., {Kroupa}, P., {Famaey}, B., \& {Fellhauer}, M. 2016,
  \mnras, 463, 1865

\bibitem[{{Duquennoy} \& {Mayor}(1991)}]{1991A&A...248..485D}
{Duquennoy}, A., \& {Mayor}, M. 1991, \aap, 500, 337

\bibitem[{{Eggleton}(1983)}]{1983ApJ...268..368E}
{Eggleton}, P.~P. 1983, \apj, 268, 368

\bibitem[{{Geha} {et~al.}(2009){Geha}, {Willman}, {Simon}, {Strigari}, {Kirby},
  {Law}, \& {Strader}}]{2009ApJ...692.1464G}
{Geha}, M., {Willman}, B., {Simon}, J.~D., {et~al.} 2009, \apj, 692, 1464

\bibitem[{{Girardi} {et~al.}(2002){Girardi}, {Bertelli}, {Bressan}, {Chiosi},
  {Groenewegen}, {Marigo}, {Salasnich}, \& {Weiss}}]{2002A&A...391..195G}
{Girardi}, L., {Bertelli}, G., {Bressan}, A., {et~al.} 2002, \aap, 391, 195

\bibitem[{{Hut} {et~al.}(1992){Hut}, {McMillan}, {Goodman}, {Mateo}, {Phinney},
  {Pryor}, {Richer}, {Verbunt}, \& {Weinberg}}]{1992PASP..104..981H}
{Hut}, P., {McMillan}, S., {Goodman}, J., {et~al.} 1992, \pasp, 104, 981

\bibitem[{{Jeans}(1919)}]{1919MNRAS..79..408J}
{Jeans}, J.~H. 1919, \mnras, 79, 408

\bibitem[{{Kirby} {et~al.}(2013){Kirby}, {Boylan-Kolchin}, {Cohen}, {Geha},
  {Bullock}, \& {Kaplinghat}}]{2013ApJ...770...16K}
{Kirby}, E.~N., {Boylan-Kolchin}, M., {Cohen}, J.~G., {et~al.} 2013, \apj, 770,
  16

\bibitem[{{Kirby} {et~al.}(2008){Kirby}, {Simon}, {Geha}, {Guhathakurta}, \&
  {Frebel}}]{2008ApJ...685L..43K}
{Kirby}, E.~N., {Simon}, J.~D., {Geha}, M., {Guhathakurta}, P., \& {Frebel}, A.
  2008, \apjl, 685, L43

\bibitem[{{Kouwenhoven} \& {de Grijs}(2008)}]{2008A&A...480..103K}
{Kouwenhoven}, M.~B.~N., \& {de Grijs}, R. 2008, \aap, 480, 103

\bibitem[{{Kroupa}(2001)}]{2001MNRAS.322..231K}
{Kroupa}, P. 2001, \mnras, 322, 231

\bibitem[{{Kroupa} \& {Burkert}(2001)}]{2001ApJ...555..945K}
{Kroupa}, P., \& {Burkert}, A. 2001, \apj, 555, 945

\bibitem[{{Massari} \& {Helmi}(2018)}]{2018A&A...620A.155M}
{Massari}, D., \& {Helmi}, A. 2018, \aap, 620, A155

\bibitem[{{Mateo}(1997)}]{1997ASPC..116..259M}
{Mateo}, M. 1997, in Astronomical Society of the Pacific Conference Series,
  Vol. 116, The Nature of Elliptical Galaxies; 2nd Stromlo Symposium, ed.
  M.~{Arnaboldi}, G.~S. {Da Costa}, \& P.~{Saha}, 259

\bibitem[{{Mateo} {et~al.}(1993){Mateo}, {Olszewski}, {Pryor}, {Welch}, \&
  {Fischer}}]{1993AJ....105..510M}
{Mateo}, M., {Olszewski}, E.~W., {Pryor}, C., {Welch}, D.~L., \& {Fischer}, P.
  1993, \aj, 105, 510

\bibitem[{{McConnachie} \& {C{\^o}t{\'e}}(2010)}]{2010ApJ...722L.209M}
{McConnachie}, A.~W., \& {C{\^o}t{\'e}}, P. 2010, \apjl, 722, L209

\bibitem[{{Meiron} \& {Kocsis}(2018)}]{2018ApJ...855...87M}
{Meiron}, Y., \& {Kocsis}, B. 2018, \apj, 855, 87

\bibitem[{{Milone} {et~al.}(2012){Milone}, {Piotto}, {Bedin}, {Aparicio},
  {Anderson}, {Sarajedini}, {Marino}, {Moretti}, {Davies}, {Chaboyer},
  {Dotter}, {Hempel}, {Mar{\'\i}n-Franch}, {Majewski}, {Paust}, {Reid},
  {Rosenberg}, \& {Siegel}}]{Milone+2012}
{Milone}, A.~P., {Piotto}, G., {Bedin}, L.~R., {et~al.} 2012, \aap, 540, A16

\bibitem[{{Minor} {et~al.}(2010){Minor}, {Martinez}, {Bullock}, {Kaplinghat},
  \& {Trainor}}]{2010ApJ...721.1142M}
{Minor}, Q.~E., {Martinez}, G., {Bullock}, J., {Kaplinghat}, M., \& {Trainor},
  R. 2010, \apj, 721, 1142

\bibitem[{{Minor} {et~al.}(2019){Minor}, {Pace}, {Marshall}, \&
  {Strigari}}]{2019MNRAS.487.2961M}
{Minor}, Q.~E., {Pace}, A.~B., {Marshall}, J.~L., \& {Strigari}, L.~E. 2019,
  \mnras, 487, 2961

\bibitem[{{Olszewski} {et~al.}(1996){Olszewski}, {Pryor}, \&
  {Armandroff}}]{1996AJ....111..750O}
{Olszewski}, E.~W., {Pryor}, C., \& {Armandroff}, T.~E. 1996, \aj, 111, 750

\bibitem[{{Rastello} {et~al.}(2020){Rastello}, {Carraro}, \&
  {Capuzzo-Dolcetta}}]{2020ApJ...896..152R}
{Rastello}, S., {Carraro}, G., \& {Capuzzo-Dolcetta}, R. 2020, \apj, 896, 152

\bibitem[{{Sepinsky} {et~al.}(2007){Sepinsky}, {Willems}, \&
  {Kalogera}}]{2007ApJ...660.1624S}
{Sepinsky}, J.~F., {Willems}, B., \& {Kalogera}, V. 2007, \apj, 660, 1624

\bibitem[{{Simon}(2019)}]{2019ARA&A..57..375S}
{Simon}, J.~D. 2019, \araa, 57, 375

\bibitem[{{Simon} \& {Geha}(2007)}]{2007ApJ...670..313S}
{Simon}, J.~D., \& {Geha}, M. 2007, \apj, 670, 313

\bibitem[{{Spencer} {et~al.}(2018){Spencer}, {Mateo}, {Olszewski}, {Walker},
  {McConnachie}, \& {Kirby}}]{2018AJ....156..257S}
{Spencer}, M.~E., {Mateo}, M., {Olszewski}, E.~W., {et~al.} 2018, \aj, 156, 257

\bibitem[{{Spencer} {et~al.}(2017){Spencer}, {Mateo}, {Walker}, {Olszewski},
  {McConnachie}, {Kirby}, \& {Koch}}]{2017AJ....153..254S}
{Spencer}, M.~E., {Mateo}, M., {Walker}, M.~G., {et~al.} 2017, \aj, 153, 254

\bibitem[{{Strigari} {et~al.}(2008){Strigari}, {Bullock}, {Kaplinghat},
  {Simon}, {Geha}, {Willman}, \& {Walker}}]{2008Natur.454.1096S}
{Strigari}, L.~E., {Bullock}, J.~S., {Kaplinghat}, M., {et~al.} 2008, \nat,
  454, 1096

\bibitem[{{Wolf} {et~al.}(2010){Wolf}, {Martinez}, {Bullock}, {Kaplinghat},
  {Geha}, {Mu{\~n}oz}, {Simon}, \& {Avedo}}]{2010MNRAS.406.1220W}
{Wolf}, J., {Martinez}, G.~D., {Bullock}, J.~S., {et~al.} 2010, \mnras, 406,
  1220

\bibitem[{{Wyse} \& {Gilmore}(2008)}]{2008IAUS..244...44W}
{Wyse}, R. F.~G., \& {Gilmore}, G. 2008, in Dark Galaxies and Lost Baryons, ed.
  J.~I. {Davies} \& M.~J. {Disney}, Vol. 244, 44--52

\end{thebibliography}
\bibliographystyle{aasjournal}

\appendix

\section{Model settings} \label{settings}

Positions and velocities of both single stars and binary centers of gravity are randomly sampled from a Plummer profile according to the algorithm proposed by \citeauthor{1974A&A....37..183A}, \citeyear{1974A&A....37..183A}.\\
Radial positions are given by
\begin{equation}
   r =  \frac{R}{\sqrt{X_{1}^{-\frac{2}{3}}-1}},
\end{equation}
and the corresponding position vector components are
\begin{equation}
\begin{aligned}
    x &= \sqrt{r^{2}-z^{2}}~\cos{(2\pi X_{3})},\\
    y &= \sqrt{r^{2}-z^{2}}~\sin{(2\pi X_{3})},\\
    z &= (1-2X_{2})~r,\\
\end{aligned}
\end{equation}
where $X_1, X_2, X_3$ are three random numbers in the interval [0,1]. We attributed the first $N_{s}$ radial position vectors to single stars ($\mathbf{r}_{s}$, with components $x_s$, ~$y_s$, ~$z_s$), and the remaining $N_b$ ones to binary centers of mass ($\mathbf{r}_{b}$, with components $x_b$, ~$y_b$, $z_b$).\\
To obtain the components of the velocity vectors, we adopted an accept-reject procedure, respecting the cut to the escape velocity at each position $\mathbf{r}$, i.e.,
\begin{equation}
    v_{esc} = \sqrt{2 U(r)},
\end{equation}

where $U(r)$ is the Plummer's potential at distance $r$ to the center. The velocity components are

\begin{equation}
\begin{aligned}
    v_x &= (1 -2X_4)~v,\\
    v_y &= \sqrt{v^2-v_x^2}~\sin{(2\pi X_5)},\\
    v_z &= \sqrt{v^2-v_x^2}~\cos{(2\pi X_5)},\\
\end{aligned}
\end{equation}

where $X_4, X_5$ are two random numbers in the interval [0,1]. Their units are, of course, those chosen for the absolute value of the velocity $v$.


Therefore, as in the case of positions, we assigned the first $N_s$ radial velocity vectors to single stars ($\mathbf{v}_{s}$, with components $v_{x,s}, ~v_{y,s}, ~v_{z,s}$), and the other $N_b$ ones to binary centers of mass ($\mathbf{v}_{b}$, with components $v_{x,b}, ~v_{y,b}, ~v_{z,b}$).\\


With regards to binary orbital parameters, i.e., the semi-major axis $a$ and the eccentricity $e$, we acted in the following way.

The generic value of $a$ is obtained as 

\begin{equation}
    a = \exp{(n_a X_a)+\ln{(a_{min})}},
\end{equation}
where $X_a$ is random number in the interval [0,1] and 
\begin{equation}
    n_a = \ln{\bigg(\frac{a_{max}}{a_{min}}\bigg)}
\end{equation}
the normalization factor.

For the eccentricity, instead, we have

\begin{equation}
    e = \sqrt{n_e X_e + e_{min}^2},
\end{equation}
where, as above, $X_e$ is a random number in the interval [0,1] and
\begin{equation}
    n_e = e_{max}^2-e_{min}^2
\end{equation}
the normalization factor.\\
Finally, we evaluated both the positions and the velocities of the $2N_b$ binary components from the center of mass reference frame and by adopting a configuration in which the secondaries are at the apocentre of the orbit of the binary system they belong to, whereas the primaries are integral with their associated center of mass.
Thus, given the apocentre radius and the orbital velocity moduli
\begin{equation}
\begin{aligned}
    r_{apo} &= a(1+e),\\
    v_{orb} &= \sqrt{\frac{Gm_b}{a}},\\
\end{aligned}
\end{equation}
we calculated the components of the corresponding vectors by means of a linear transformation to map random numbers from the interval [0,1] to the interval [-1,1]. In this way, the position and velocity vectors of primaries result
\begin{equation}
\begin{aligned}
    \mathbf{r}_{1} &= \mathbf{r}_{b} + \frac{m_2}{m_b} ~\mathbf{r}_{apo},\\
    \mathbf{v}_{1} &= \mathbf{v}_b + \frac{m_2}{m_b} ~\mathbf{v}_{orb},\\
\end{aligned}
\end{equation}
whereas those of secondaries are
\begin{equation}
\begin{aligned}
    \mathbf{r}_{2} &= \mathbf{r}_{b} - \frac{m_1}{m_b} ~\mathbf{r}_{apo},\\
    \mathbf{v}_{2} &= \mathbf{v}_{b} - \frac{m_1}{m_b} ~\mathbf{v}_{orb}.\\
\end{aligned}
\end{equation}

\section{Velocity dispersion} \label{sigma}

\subsection{Velocity dispersion in the absence of RLOF}

Following a scheme similar to that outlined in \citeauthor{2020ApJ...896..152R}, \citeyear{2020ApJ...896..152R}, we examined various possible ways to estimate the system velocity dispersion:
\begin{enumerate}
    \item by considering all the stars as if they were single and resolved, so that each binary component counts as one star:
    \begin{equation} \label{sigma_tot}
        \sigma_{tot} = \sqrt{\frac{\sum\limits_{i=1}^{N} (\mathbf{v}_{i} - \langle{v}\rangle)^2}{N}},
    \end{equation}
    where
   \begin{equation}
        \langle{v}\rangle = \frac{\sum\limits_{i=1}^{N}\mathbf{v}_{i}}{N};
    \end{equation}
    \item by distinguishing the contribution of single stars from that of binaries, which are represented by their own center of mass:
    \begin{equation} \label{sigma_sb}
        \sigma_{sb}^2 = \frac{\sum\limits_{i=1}^{N_s+N_b}(\mathbf{v}_{i} - \langle{v}\rangle)^2}{N_s+N_b},
    \end{equation}
    with
    \begin{equation}
        \langle{v}\rangle = \frac{\sum\limits_{i=1}^{N_s+N_b}\mathbf{v}_{i}}{N_s+N_b};
    \end{equation}
    \item by neglecting the presence of binary stars, thus accounting for the contribution of single stars only. As a consequence,
    \begin{equation} \label{sigma_s}
        \sigma_s = \sqrt{\frac{\sum\limits_{i=1}^{N_s}(\mathbf{v}_{i} - \langle{v}\rangle)^2}{N_s}};
    \end{equation}
    \item By weighting the velocity of both single and binary components by their luminosity, according to the evolutionary type. Therefore, this way of estimating the velocity dispersion differs from the first one only in the average of stellar velocities
    \begin{equation}
        \langle{v}\rangle = \frac{\sum\limits_{i=1}^{N} L_i\mathbf{v}_{i}}{\sum\limits_{i=1}^{N}L_i},
    \end{equation}
    so that
    \begin{equation} \label{sigma_tot_lum}
        \sigma_{tot,lum} = \sqrt{\frac{\sum\limits_{i=1}^{N} L_i(\mathbf{v}_{i}-\langle{v}\rangle)^2}{\sum\limits_{i=1}^{N}L_i}}.
    \end{equation}
    In particular, the luminosity of both MS and RGB stars has been determined by fitting an isochrone of 13 Gyrs from the Padua stellar and evolutionary tracks and isochrones database (\citeauthor{2002A&A...391..195G}, \citeyear{2002A&A...391..195G});
    \item by weighting the velocity of single stars, only, by their luminosity according to the evolutionary type, i.e.,
    \begin{equation}
        \langle{v}\rangle = \frac{\sum\limits_{i=1}^{N_s} L_i\mathbf{v}_{i}}{\sum\limits_{i=1}^{N_s}L_i},
    \end{equation}
    which implies that
    \begin{equation} \label{sigma_s_lum}
        \sigma_{s,lum} = \sqrt{\frac{\sum\limits_{i=1}^{N_s}L_i(\mathbf{v}_{i}-\langle{v}\rangle)^2}{\sum\limits_{i=1}^{N_s}L_i}}.
    \end{equation}
    \end{enumerate}

\subsection{Velocity dispersion in the presence of RLOF}

When accounting for RLOF, we made a distinction between the accepted and rejected binaries as far as the calculation of the velocity dispersion is concerned: the former contribute with their components' orbital motion, whereas the latter with the center of mass velocity. Hence, formula \ref{sigma_tot} becomes

\begin{equation} \label{sigma_tot_RLOF}
    \sigma_{tot} = \sqrt{\frac{\sum\limits_{i=1}^{N_s} (\mathbf{v}_{i} - \langle{v}\rangle)^2 + \sum\limits_{i=1}^{2N_{b,acc}}(\mathbf{v}_{i} - \langle{v}\rangle)^2 + \sum\limits_{i=1}^{N_{b,rej}} (\mathbf{v}_{i} - \langle{v}\rangle)^2}{N_s+2N_{b,acc}+N_{b,rej}}},
\end{equation}

where

\begin{equation}
    \langle{v}\rangle = \frac{\sum\limits_{i=1}^{N_s}\mathbf{v}_{i} + \sum\limits_{i=1}^{2N_{b,acc}}\mathbf{v}_{i} + \sum\limits_{i=1}^{N_{b,rej}}\mathbf{v}_{i}}{N}.
\end{equation}

Instead, formula \ref{sigma_tot_lum} takes the form

\begin{equation} \label{sigma_tot_lum_RLOF}
    \sigma_{tot,lum} = \sqrt{\frac{\sum\limits_{i=1}^{N_s} L_i(\mathbf{v}_{i}-\langle{v}\rangle)^2 + \sum\limits_{i=1}^{2N_{b,acc}} L_i(\mathbf{v}_{i}-\langle{v}\rangle)^2 + \sum\limits_{i=1}^{N_{b,rej}} (L_{1,i}+L_{2,i})(\mathbf{v}_{i}-\langle{v}\rangle)^2 }{\sum\limits_{i=1}^{N_s}L_i + \sum\limits_{i=1}^{2N_{b,acc}}L_i + \sum\limits_{i=1}^{N_{b,rej}} (L_{1,i}+L_{2,i})}},
\end{equation}

with
    
\begin{equation}
    \langle{v}\rangle = \frac{\sum\limits_{i=1}^{N_s} L_i\mathbf{v}_{i} + \sum\limits_{i=1}^{2N_{b,acc}} L_i\mathbf{v}_{i} + \sum\limits_{i=1}^{N_{b,rej}} (L_{1,i}+L_{2,i})\mathbf{v}_{i}}{\sum\limits_{i=1}^{N_s}L_i + \sum\limits_{i=1}^{2N_{b,acc}}L_i + \sum\limits_{i=1}^{N_{b,rej}} (L_{1,i}+L_{2,i})},
\end{equation}

where the velocity of rejected binaries is weighted by the sum of their respective components' luminosities.

\end{document}